# Forging a Developed India: Growth Imperatives, Fiscal Sustainability, and Multilateral Partnerships for Viksit Bharat 2047

Supriya Sanjay Nikam

Fellow B, Mumbai School of Economics and Public Policy, University of Mumbai, Mumbai

supriyanikam2002@gmail.com

Satyanarayan Kothe

Professor, Mumbai School of Economics and Public Policy, University of Mumbai, Mumbai

kothesk@gmail.com

https://orcid.org/0000-0002-6496-0129

## Abstract

This paper examines the fiscal and macroeconomic strategies essential for India's transition to a high-income economy by 2047, aligning with the vision of Viksit Bharat. A sustainable annual GDP growth rate of 7–8 percent is projected as necessary to achieve this milestone while maintaining fiscal prudence through a targeted deficit threshold below 3.5 percent of GDP. The study underscores the role of disciplined fiscal management in financing critical public investments in infrastructure, human capital development and technological innovation. Given constraints on domestic resource mobilization, the paper highlights the importance of multilateral financial institutions—including the World Bank, IMF and ADB—in expanding India's fiscal space through concessional financing, technical cooperation, and risk-sharing mechanisms. Using econometric modeling and scenario analysis, the research identifies key policy interventions in infrastructure, healthcare, education and sustainable energy that can maximize growth while ensuring fiscal sustainability. Policy recommendations include enhancing tax buoyancy, rationalizing expenditure, optimizing public-private partnerships (PPPs) and strengthening fiscal responsibility frameworks. The findings suggest that a calibrated approach to growth, prudent fiscal management and strategic international collaborations are critical to achieving India's long-term economic aspirations.

**Introduction:**

India aspires to be developed economy by 2047. There are numerous challenges to achieve the milestone. The country needs to strategies on macroeconomic policies including fiscal prudence on the path of sustainability. This paper examines the fiscal and macroeconomic strategies essential for India's transition to a high-income economy by 2047, aligning with the vision of Viksit Bharat. A sustainable annual GDP growth rate of 7–8 percent is projected as necessary to achieve this milestone while maintaining fiscal prudence through a targeted deficit threshold below 3.5 percent of GDP. The study underscores the role of disciplined fiscal management in financing critical public investments in infrastructure, human capital development and technological innovation. Given constraints on domestic resource mobilization, the paper highlights the importance of multilateral financial institutions—including the World Bank, IMF and ADB—in expanding India's fiscal space through concessional financing, technical cooperation, and risk-sharing mechanisms. Using econometric modeling and scenario analysis, the research identifies key policy interventions in infrastructure, healthcare, education and sustainable energy that can maximize growth while ensuring fiscal sustainability. Policy recommendations include enhancing tax buoyancy, rationalizing expenditure, optimizing public-private partnerships (PPPs) and strengthening fiscal responsibility frameworks. The findings suggest that a calibrated approach to growth, prudent fiscal management and strategic international collaborations are critical to achieving India's long-term economic aspirations.

**Review of Literature:**

Various studies tried to analyze the linkage between fiscal deficit and economic growth. Results are widely skewed, some studies found unidirectional causality whereas others found bidirectional causality. Both in the long run and the short run fiscal deficit and revenue deficit have an adverse effect on economic growth (Mohanty, 2018 and 2020). Mohanty (2020) study found that fiscal deficit influences economic growth both directly and indirectly through routes of investment, interest rate, current account deficit and composition of government expenditure. Kumar and Kumar (2021) study found that there was unidirectional causality from fiscal deficit to GDP growth, while Mohanty (2020) study found that there exists a bi-directional relationship between fiscal deficit and economic growth in the long run. Kumar and

Kumar (2021) study showed that in the long run, fiscal deficit had a significant negative impact on economic growth as a one percent increase in fiscal deficit demoted the GDP growth rate by 0.075 percent. In contrast, in the short run, the effect was also found negative, but it was significant with only one lag (Kumar and Kumar, 2021).

Studies further try to evaluate the impact of the FRBM Act in managing fiscal deficits. Sethi et.al. (2019) found that the adverse impact of fiscal deficit on economic growth is almost the same in both pre and post-FRBM act periods, whereas Mohanty (2020) study revealed that implementation of the FRBM Act has influenced and weakened the relationship between fiscal deficit and economic growth in India. The Government should contain the fiscal deficit and should try to achieve the target set by the FRBM Act (Mohanty, 2018). India was able to achieve the target of 3% of GDP only once in 2007-08 (Mohanty, 2020).

The method of deficit financing and the existing public debt stock influence the relationship between fiscal deficits and economic growth. Taxes and grants have relatively clear effects on growth, but the impact of deficits is more nuanced. Deficits can support growth when financed through limited seigniorage, whereas reliance on domestic debt tends to be growth-constraining. External borrowing at market rates introduces both short-term (flow) and long-term (stock) effects that may work in opposing directions. Furthermore, the relationship is likely to exhibit two forms of non-linearity: one linked to the size of the deficit and another arising from interactions between the deficit and public debt levels (Adam and Bevan, 2004). Avila (2011) finds that fiscal deficits, through the macroeconomic volatility they generate—particularly in relative prices—serve as a structural constraint on per capita income growth in Argentina over the long term (1915–2006). Tung (2028) highlighted that fiscal deficit has a harmful effect on economic growth in the long run in Vietnam and the study got the coefficient of Fiscal Deficit as -3.34.

A significant share of resources generated through fiscal deficit is used for relatively unproductive purposes such as interest payment and other committed expenditures (Mohanty, 2020). The study suggested that the government should reduce non-productive expenditure, manage available resources efficiently, and generate new revenue sources to reduce dependency on borrowing (Kumar and Kumar, 2021).

**International Organization:**

International organizations like the International Fund for Agricultural Development (IFAD) have been instrumental in improving food security through targeted interventions. For instance, IFAD's programs focus on enhancing small farmers' access to finance, improving land and water management, and increasing resilience to climate change (Kozhukhova, 2016 and Albert & Deekor, 2014).

The Japan International Cooperation Agency (JICA) has implemented agricultural development projects in Cameroon, leading to increased crop yields, improved income, and enhanced well-being among beneficiaries. Such projects underscore the importance of international cooperation in addressing food insecurity and poverty (Bamenju et al., 2022).

Improved rural roads enable farmers to access markets more efficiently, increasing their income and productivity. International aid programs, such as those supported by the World Bank, have prioritized rural road rehabilitation to enhance market access and economic opportunities (No. 52531. International Development Association and Congo, 2022 and Cleaver, 1997).

IFAD has implemented programs that focus on training women, men, and youth in skills acquisition and leadership development. These initiatives have empowered rural communities to take charge of their development and improve their income-earning capabilities (Albert & Deekor, 2014 and Harry, 2016).

International Fund for Agricultural Development (IFAD) projects in Nigeria and Cameroon have led to significant improvements in rural infrastructure, including schools, water boreholes, and training programs. These interventions have enhanced human capacity and income levels (M., n.d. and Albert & Deekor, 2014).

Aid programs often encourage diversification of livelihood activities, reducing dependency on single crops and enhancing resilience against economic shocks (Muluh et al., 2019 and Bamenju et al., 2022).

Swedish development aid has successfully reduced poverty in regions like Sub-Saharan Africa by focusing on sustainable livelihood approaches, including land certification and gender equality initiatives (Arefaine et al., 2015). Similarly, IFAD interventions in Nigeria have improved the standard of living for internally displaced farmers (Samuel et al., 2022).

Challenges such as corruption, weak institutions, and misallocation of funds have hindered the effectiveness of aid in some regions. For instance, in Nepal, the functional use of aid in agriculture declined despite increased inflows (Bhandari, 2024 and Ssozi et al., 2017).

**Education**

India's experience with foreign aid for education highlights the importance of aligning aid with national priorities. While donors influenced policy implementation, India maintained control over its educational goals, ensuring that external resources were used to complement domestic initiatives (Colclough & Webb, 2010 and Tilak, 2008).

The SSA program, supported by foreign aid, significantly improved access to elementary education, particularly for disadvantaged groups. The program's success was attributed to its alignment with India's national policy and the harmonization of donor practices (Ward, 2011).

India has emerged as a key player in global education, with foreign aid fostering international collaborations and positioning India as a potential global educational hub (Khare, 2015 and Oriel, 2023).

The effectiveness of foreign aid is highly dependent on the governance and political context of the recipient country. Countries with stable governance structures tend to benefit more from educational aid (Turrent, 2016).

Agricultural official development assistance (ODA) can facilitate foreign direct investment in agriculture, fishery, and forestry sectors, thereby enhancing the overall investment climate in these sectors (Tian, 2023).

**<u>Data and Methodology:</u>**

Box Jenkins (1970) introduced a three-step method for appropriate model selection for estimating and forecasting univariate models. The three steps are identification, estimation, and diagnostics. The identification step comprises checking stationarity and determination of Autoregressive, difference and moving average components. If variables are stationary then we use ARMA models and if they are non-stationary, we use ARIMA models.

Variable per Capita GNI is taken from the World Bank Database. World Bank used the Atlas method to calculate GNI per capita (current US $). As data was available from 1962 to 2023, the entire estimation is based on the available data.

Figure 1: Per Capita GNI (Current US $).

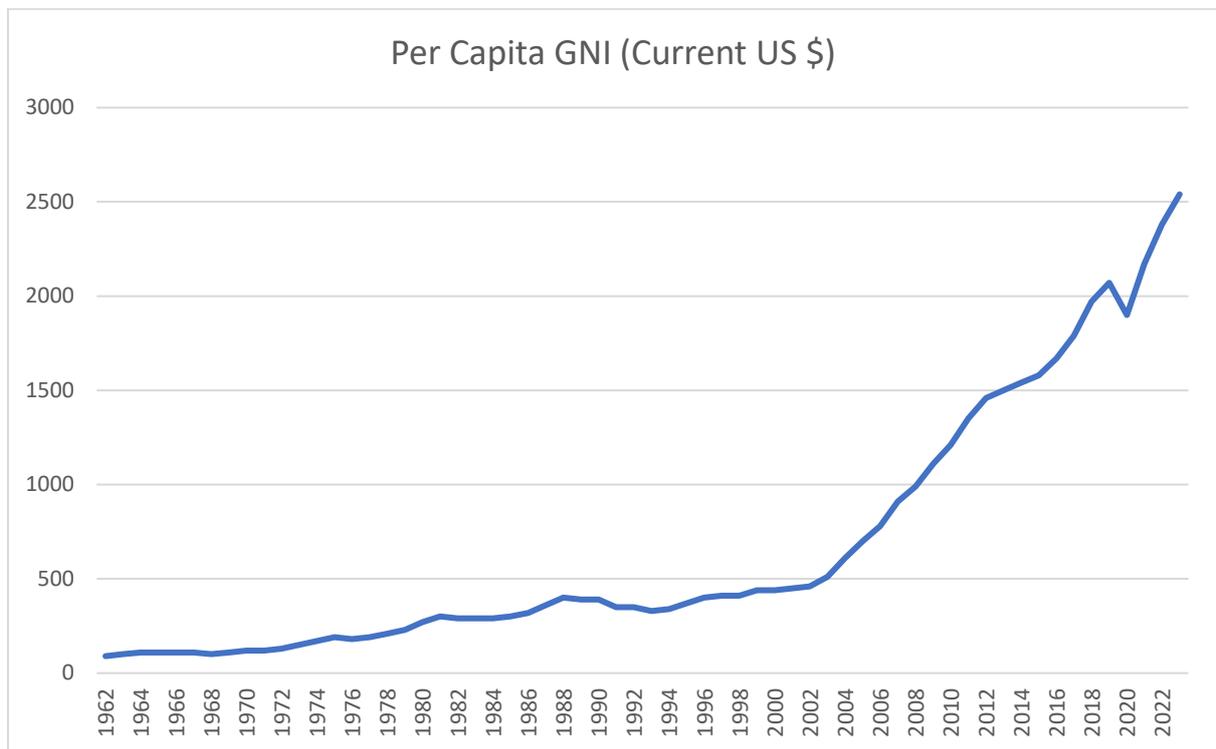

Source: World Bank Database

Figure 2: Partial Autocorrelation Function (PACF) for level form per capita GNI

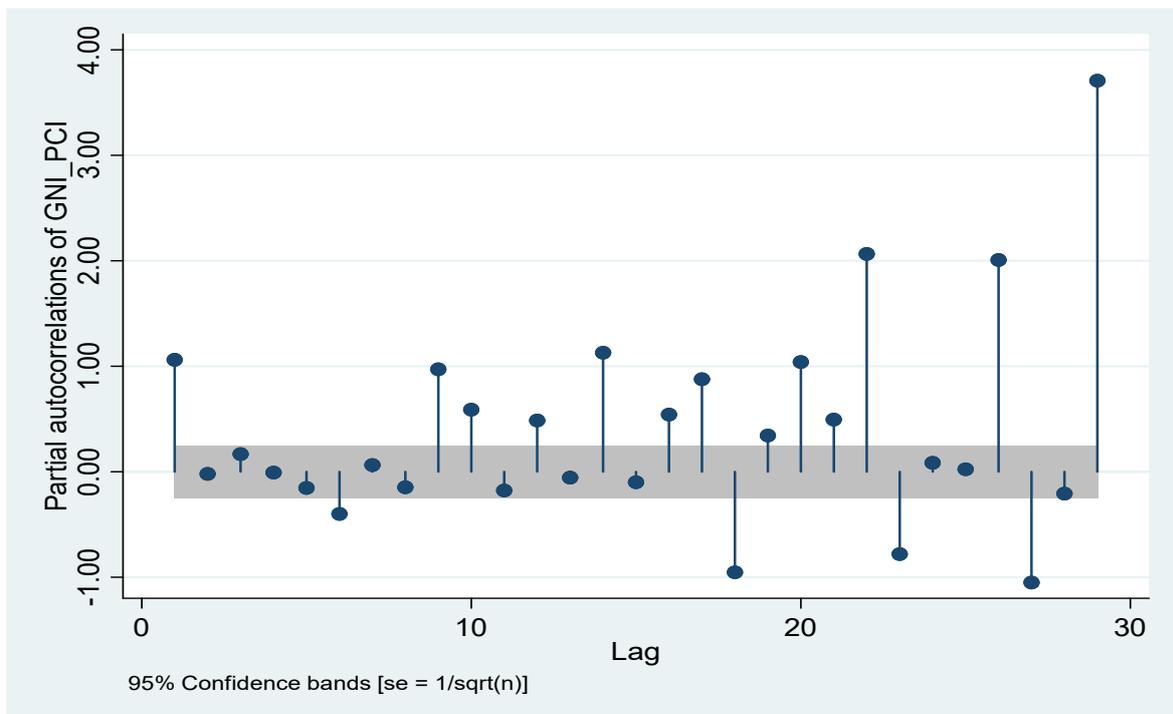

Source: Author's computation.

As it is seen from the figure Per Capita GNI is not stationary as it is showing a positive trend.

For the identification of Autoregressive and moving average components Partial Autocorrelation Function (PACF) and Autocorrelation Function (ACF) are constructed. They are as follows

Figure 3: Autocorrelation Function (ACF) for level form per capita GNI

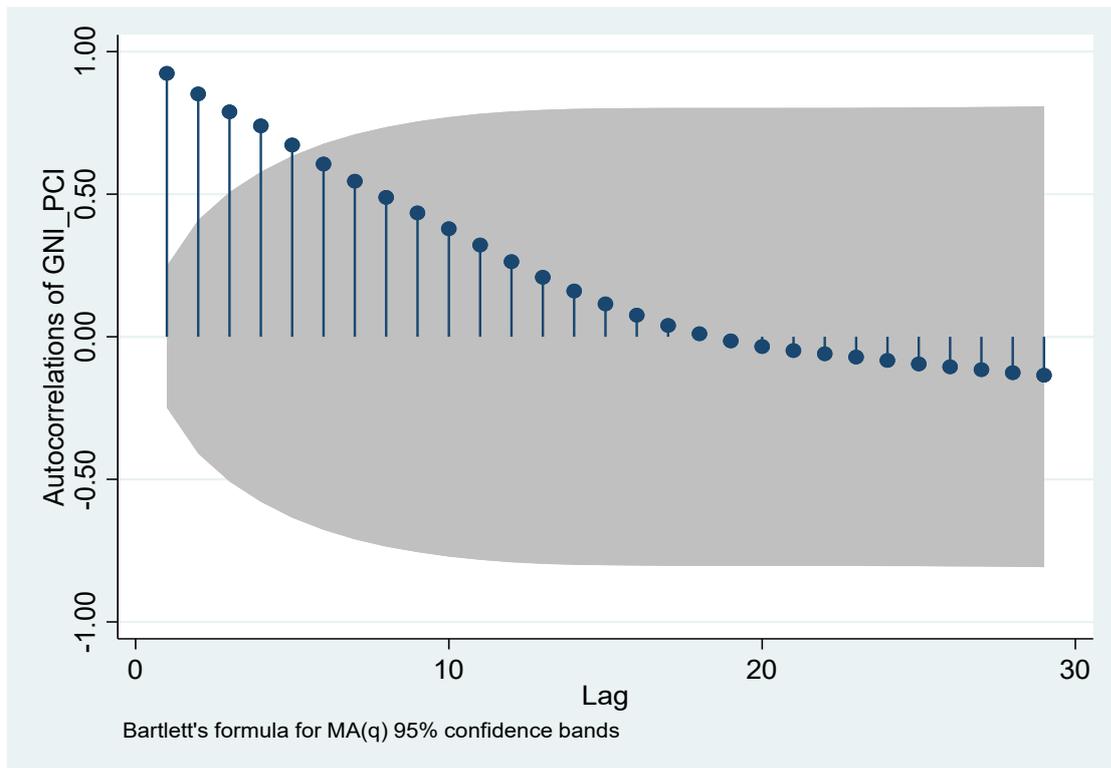

Source: Author's computation.

ACF is showing gradual decay after the 1$^{st}$ lag. Figures 2 and 3 show that Per Capita GNI is not stationary. Further, it is complemented by Augmented Dicky Fuller (ADF) and Phillips Perron (PP) tests, which are tests for stationarity.

Table 1: Augmented Dickey-Fuller test for unit root

| Variables | Test Statistics Z (t) | P-Value | 1 % Critical Value | 5 % Critical Value | 10 % Critical Value |
|---|---|---|---|---|---|
| $Per\ Capita\ GNI$ | 1.796 | 1.0000 | -4.126 | -3.489 | -3.173 |
| $Per\ Capita\ GNI_{t-1}$ | -6.788*** | 0.0000 | -4.128 | -3.49 | -3.174 |

Note: *p<0.01, **p<0.05, ***p < 0.001

Source: Author's computation.

Results of Augmented Dicky Fuller unit root test showed that for level form per capita GNI, test statistics i.e. Z(t) lie beyond the confidence interval, and the P-value is also greater than 0.05. Hence, we failed to reject the null hypothesis (time series data is non-stationary). Further, the test is performed on the first difference per capita GNI and the result showed that the differenced Per Capita GNI is stationary at I (1) as test statists lie in the confidence interval and the P-value is less than 0.05. Hence, we are rejecting the null of time series data is non-stationary.

Table 2: Phillips-Perron test for unit root

| Variables | Test Statistics | | P-Value | 1 % Critical Value | 5 % Critical Value | 10 % Critical Value |
|---|---|---|---|---|---|---|
| *Per Capita GNI* | Z(rho) | 2.619 | 1.0000 | -26.074 | -19.998 | -16.954 |
| | Z(t) | 2.025 | | -4.126 | -3.489 | -3.173 |
| *Per Capita GNI$_{t-1}$* | Z(rho) | -51.40 | 0.0000 | -26.04 | -19.98 | -16.94 |
| | Z(t) | -6.751 | | -4.128 | -3.49 | -3.174 |

Note: *p<0.01, **p<0.05, ***p < 0.001

Source: Author's computation.

Results of the Phillips-perron unit root test showed that Per Capita GNI at its level form is non-stationary as test statistics lie outside the confidence interval and the p-value is greater than 0.05. Hence, failed to reject the null of non-stationarity. The same test is performed for difference Per Capita GNI, and found that even at 1% significance we are rejecting the null of non-stationarity and accepting the alternative of stationary time series.

In short, the results of Augmented Dicky Fuller and Phillips Perron's test indicated that Per Capita GNI is non-stationary at the level form but it became stationary at first difference.

Figure 4: First difference Per Capita GNI

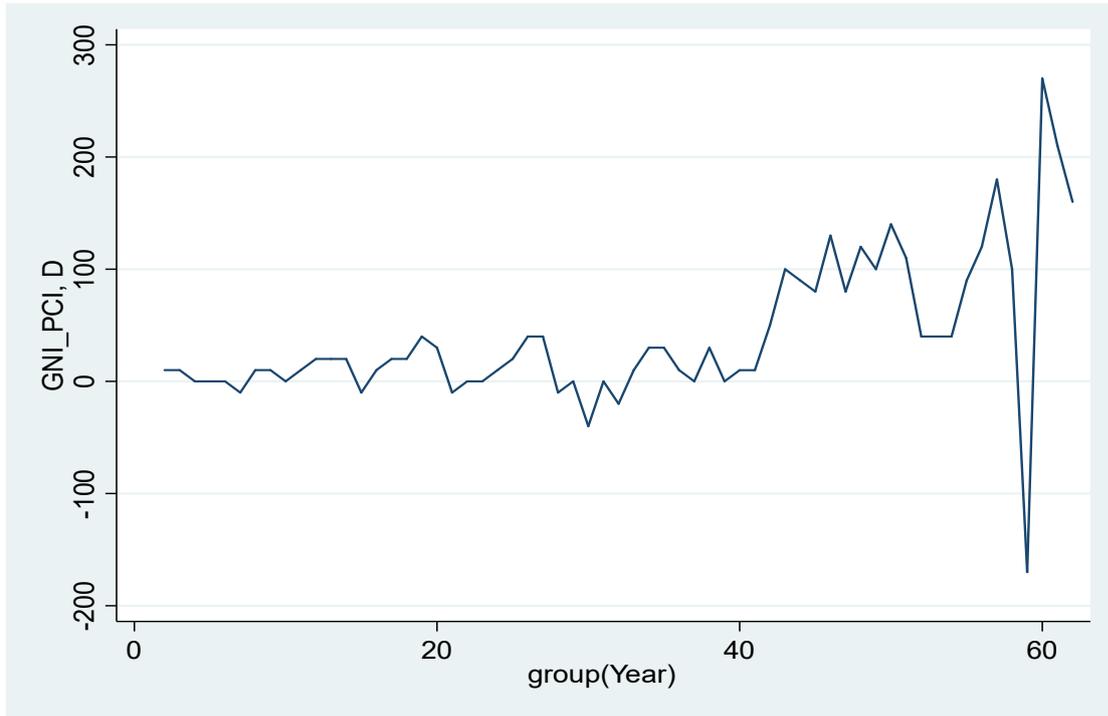

Source: Author's computation

Per Capita GNI on an average showing stationarity after taking the first difference. Even though there are some fluctuations, there is no clear upward or downward trend. This suggests that taking the first difference is appropriate for stationarity.

For the model selection and forecasting, the entire period from which data is available, i.e., 1962-2023, was considered, and then the period after the LPG policy, i.e., 1991-2023, was considered.

To choose the best-fitted ARIMA models, the AIC and BIC of various ARIMA models were found with the help of Python. Different models with their AIC and BIC for the entire and sub-period are given in Tables 3 and 4.

Table 3: ARIMA models with their AIC and BIC criteria for the entire period (1962-2023)

| ARIMA Model | AIC | BIC |
| --- | --- | --- |
| (0, 0, 0) | 986.0041485 | 990.2584173 |
| (0, 0, 1) | 913.2217589 | 919.6031621 |
| (0, 0, 2) | 854.042589 | 862.5511265 |
| (0, 0, 3) | 813.7963266 | 824.4319985 |
| (0, 1, 0) | 705.0937123 | 707.2045862 |
| (0, 1, 1) | 693.2147225 | 697.4364702 |
| (0, 1, 2) | 691.3648368 | 697.6974584 |
| (0, 1, 3) | 690.9013677 | 699.3448632 |
| (0, 2, 0) | 687.1202516 | 689.2145962 |
| (0, 2, 1) | 660.5542231 | 664.7429122 |
| (0, 2, 2) | 662.5113531 | 668.7943868 |
| (0, 2, 3) | 662.1621533 | 670.5395316 |
| (1, 0, 0) | 726.3096091 | 732.6910122 |
| (1, 0, 1) | 714.4377279 | 722.9462655 |
| (1, 0, 2) | 712.5293611 | 723.165033 |
| (1, 0, 3) | 712.2037529 | 724.9665592 |
| (1, 1, 0) | 686.2089811 | 690.4307288 |
| (1, 1, 1) | 673.943988 | 680.2766096 |
| (1, 1, 2) | 675.8795184 | 684.3230138 |
| (1, 1, 3) | 675.6144744 | 686.1688437 |
| (1, 2, 0) | 678.123717 | 682.3124062 |
| (1, 2, 1) | 662.5281896 | 668.8112233 |
| (1, 2, 2) | 663.4136571 | 671.7910353 |
| (1, 2, 3) | 663.9710578 | 674.4427806 |
| (2, 0, 0) | 707.4159853 | 715.9245228 |
| (2, 0, 1) | 695.3760481 | 706.01172 |
| (2, 0, 2) | 697.8506947 | 710.613501 |
| (2, 0, 3) | 698.5746514 | 713.4645921 |
| (2, 1, 0) | 685.3222747 | 691.6548963 |
| (2, 1, 1) | 675.9044565 | 684.347952 |
| (2, 1, 2) | 677.2635932 | 687.8179625 |
| (2, 1, 3) | 677.3711873 | 690.0364305 |
| (2, 2, 0) | 669.075736 | 675.3587697 |
| (2, 2, 1) | 662.4979844 | 670.8753626 |
| (2, 2, 2) | 662.9711399 | 673.4428627 |
| (2, 2, 3) | 663.1048805 | 675.6709478 |
| (3, 0, 0) | 706.6041569 | 717.2398288 |
| (3, 0, 1) | 711.2818362 | 724.0446426 |
| (3, 0, 2) | 716.7019449 | 731.5918856 |
| (3, 0, 3) | 715.6753385 | 732.6924135 |
| (3, 1, 0) | 681.473148 | 689.9166434 |
| (3, 1, 1) | 675.932309 | 686.4866783 |
| (3, 1, 2) | 676.2567044 | 688.9219476 |
| (3, 1, 3) | 679.2140024 | 693.9901195 |
| (3, 2, 0) | 664.7645997 | 673.1419779 |
| (3, 2, 1) | 662.6699421 | 673.1416649 |
| (3, 2, 2) | 664.0370081 | 676.6030755 |
| (3, 2, 3) | 664.9473822 | 679.6077941 |

Source: Author's computation.

Table 4: ARIMA models with their AIC and BIC criteria for the sub-period (1991-2023)

| ARIMA Model | AIC | BIC |
|---|---|---|
| (0, 0, 0) | 528.9153258 | 531.9083409 |
| (0, 0, 1) | 493.3875627 | 497.8770854 |
| (0, 0, 2) | 465.1563218 | 471.142352 |
| (0, 0, 3) | 448.0573973 | 455.5399351 |
| (0, 1, 0) | 390.5109638 | 391.9766997 |
| (0, 1, 1) | 385.2439541 | 388.1754259 |
| (0, 1, 2) | 385.3381184 | 389.7353261 |
| (0, 1, 3) | 386.0852624 | 391.948206 |
| (0, 2, 0) | 375.3745081 | 376.8084953 |
| **(0, 2, 1)** | **362.5970197** | **365.4649942** |
| (0, 2, 2) | 364.592793 | 368.8947546 |
| (0, 2, 3) | 364.8998758 | 370.6358246 |
| (1, 0, 0) | 411.3400659 | 415.8295886 |
| (1, 0, 1) | 406.0753107 | 412.061341 |
| (1, 0, 2) | 405.9765427 | 413.4590805 |
| (1, 0, 3) | 406.9671579 | 415.9462032 |
| (1, 1, 0) | 381.6493908 | 384.5808626 |
| (1, 1, 1) | 376.1722874 | 380.5694951 |
| (1, 1, 2) | 378.1580377 | 384.0209813 |
| (1, 1, 3) | 378.3254129 | 385.6540924 |
| (1, 2, 0) | 371.7080036 | 374.575978 |
| (1, 2, 1) | 364.594686 | 368.8966476 |
| (1, 2, 2) | 364.9937415 | 370.7296903 |
| (1, 2, 3) | 366.7195299 | 373.889466 |
| (2, 0, 0) | 402.2783777 | 408.2644079 |
| (2, 0, 1) | 392.6526391 | 400.1351769 |
| (2, 0, 2) | 398.7266002 | 407.7056456 |
| (2, 0, 3) | 399.7757893 | 410.2513422 |
| (2, 1, 0) | 382.152661 | 386.5498688 |
| (2, 1, 1) | 378.164246 | 384.0271896 |
| (2, 1, 2) | 379.8280661 | 387.1567456 |
| (2, 1, 3) | 379.561181 | 388.3555964 |
| (2, 2, 0) | 367.5459994 | 371.8479611 |
| (2, 2, 1) | 365.2619783 | 370.9979271 |
| (2, 2, 2) | 363.1933227 | 370.3632587 |
| (2, 2, 3) | 363.1621069 | 371.7660301 |
| (3, 0, 0) | 402.9573122 | 410.4398501 |
| (3, 0, 1) | 406.3001891 | 415.2792344 |
| (3, 0, 2) | 410.9215739 | 421.3971268 |
| (3, 0, 3) | 402.717281 | 414.6893415 |
| (3, 1, 0) | 380.8116751 | 386.6746187 |
| (3, 1, 1) | 378.8891575 | 386.217837 |
| (3, 1, 2) | 376.0487025 | 384.8431179 |
| (3, 1, 3) | 383.352171 | 393.6123224 |
| (3, 2, 0) | 365.9148058 | 371.6507546 |
| (3, 2, 1) | 365.8862412 | 373.0561772 |
| (3, 2, 2) | 369.0335561 | 377.6374793 |
| (3, 2, 3) | 366.7615338 | 376.7994442 |

Source: Author's computation.

The per Capita GNI forecast for the entire and sub-period is as follows.

Table 5: Forecast based on entire period (1962-2023)

| Year | Forecast |
|---|---|
| 2024 | 2663.01165 |
| 2025 | 2786.0233 |
| 2026 | 2909.034951 |
| 2027 | 3032.046601 |
| 2028 | 3155.058251 |
| 2029 | 3278.069901 |
| 2030 | 3401.081551 |
| 2031 | 3524.093202 |
| 2032 | 3647.104852 |
| 2033 | 3770.116502 |
| 2034 | 3893.128152 |
| 2035 | 4016.139803 |
| 2036 | 4139.151453 |
| 2037 | 4262.163103 |
| 2038 | 4385.174753 |
| 2039 | 4508.186403 |
| 2040 | 4631.198054 |
| 2041 | 4754.209704 |
| 2042 | 4877.221354 |
| 2043 | 5000.233004 |
| 2044 | 5123.244654 |
| 2045 | 5246.256305 |
| 2046 | 5369.267955 |
| 2047 | 5492.279605 |

Source: Author's computation.

Figure 5: Forecast based on entire period (1962-2023)

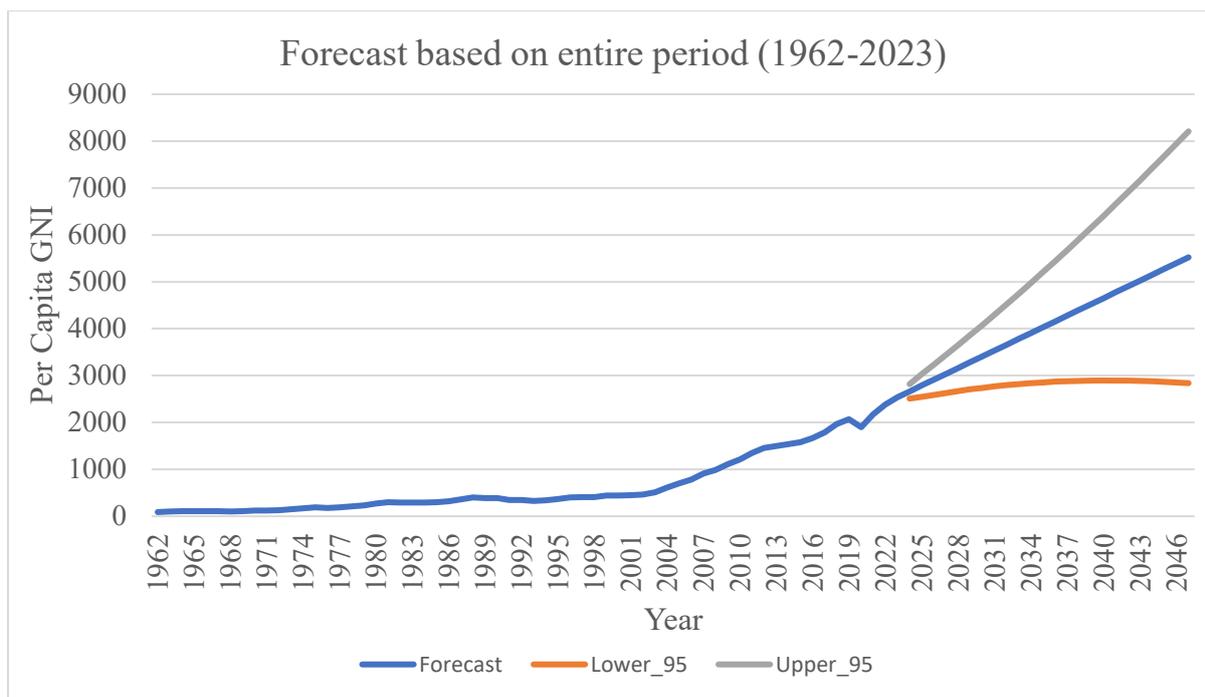

Source: Author's computation.

Table 6: Forecast based on sub-period (1991-2023)

| Year | Forecast |
|---|---|
| 2024 | 2664.23449 |
| 2025 | 2788.468979 |
| 2026 | 2912.703469 |
| 2027 | 3036.937959 |
| 2028 | 3161.172448 |
| 2029 | 3285.406938 |
| 2030 | 3409.641428 |
| 2031 | 3533.875917 |
| 2032 | 3658.110407 |
| 2033 | 3782.344896 |
| 2034 | 3906.579386 |
| 2035 | 4030.813876 |
| 2036 | 4155.048365 |
| 2037 | 4279.282855 |
| 2038 | 4403.517345 |
| 2039 | 4527.751834 |
| 2040 | 4651.986324 |
| 2041 | 4776.220814 |
| 2042 | 4900.455303 |
| 2043 | 5024.689793 |
| 2044 | 5148.924283 |
| 2045 | 5273.158772 |
| 2046 | 5397.393262 |
| 2047 | 5521.627752 |

Source: Author's computation.

Figure 6: Forecast based on sub-period (1991-2023)

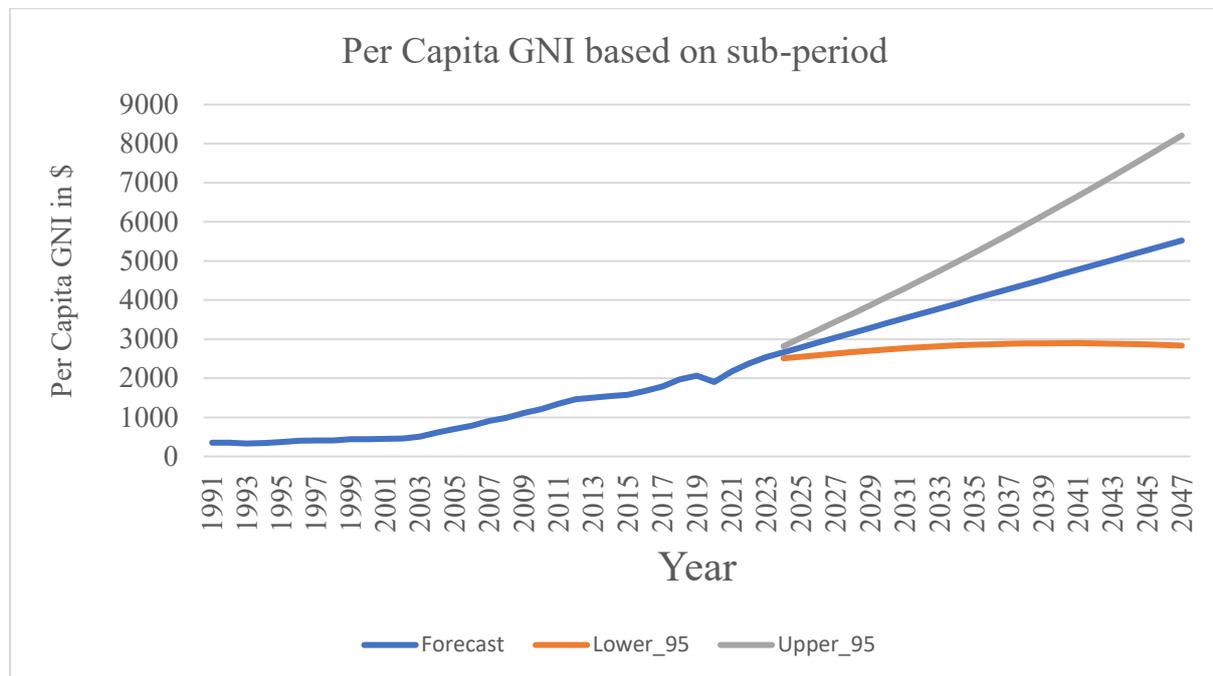

Source: Author's computation.

As per the forecasted value of per capita GNI, the annual average growth rate is 3%. If India wants to be in the developed category status India's per capita GNI has to grow by an annual average of 7%. The growth rate is also the same for the sub-period. As per the World Bank's calculation of GNI per capita based on the World Bank Atlas Method, lower-middle-income economies are those with a GNI per capita between $1,146 and $4,515; upper-middle-income economies are those with a GNI per capita between $4,516 and $14,005; high-income economies are those with more than a GNI per capita of $14,005.

Data for Gross Fiscal Deficit and Gross Domestic Product (in Rs. Crores) is extracted from the RBI dataset. For Gross Fiscal Deficit data was available from 1971 to 2025, whereas for Gross Domestic Product data is available from 1951 to 2025.

Figure 7: Gross Fiscal Deficit (in Rs. Crores)

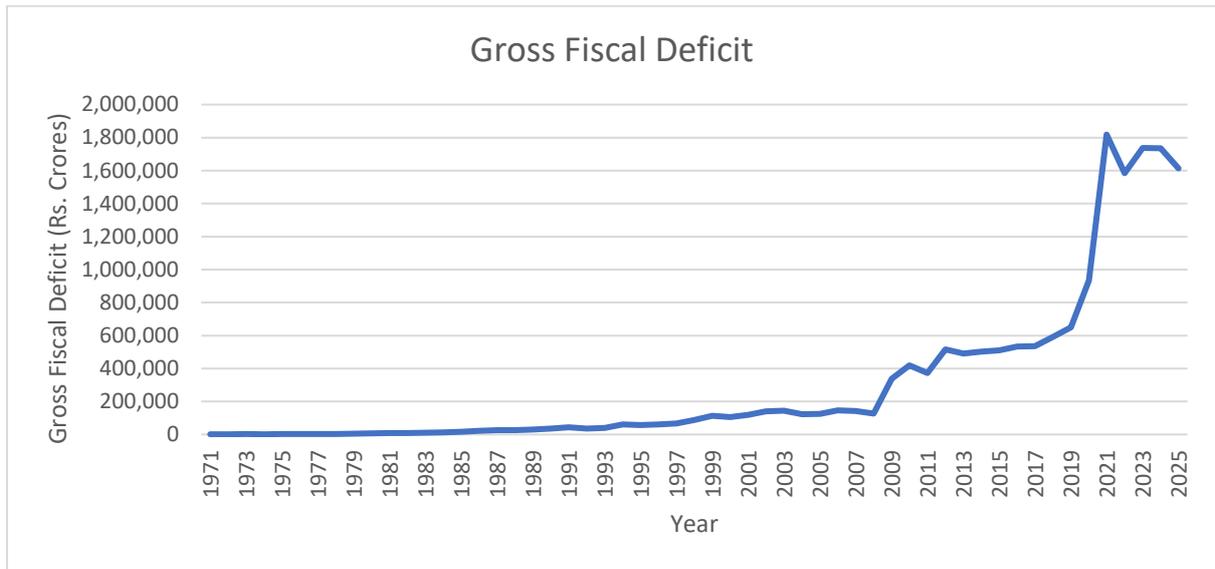

Source: RBI dataset.

For the identification of Autoregressive and moving average components Partial Autocorrelation Function (PACF) and Autocorrelation Function (ACF) are constructed. They are as follows

Figure 8: Partial Autocorrelation Function (PACF) for level form Gross Fiscal Deficit

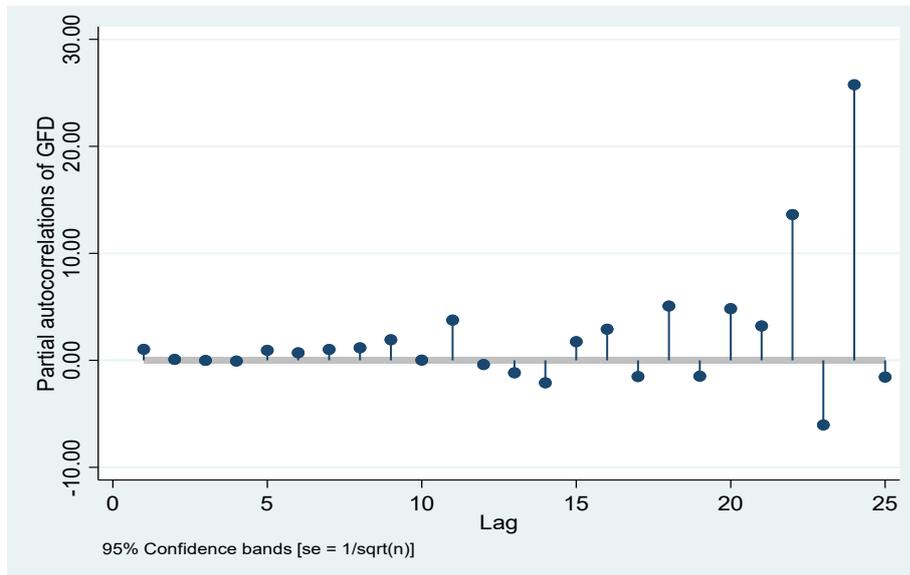

Source: Author's computation.

Figure 9: Autocorrelation Function (ACF) for level form Gross Fiscal Deficit

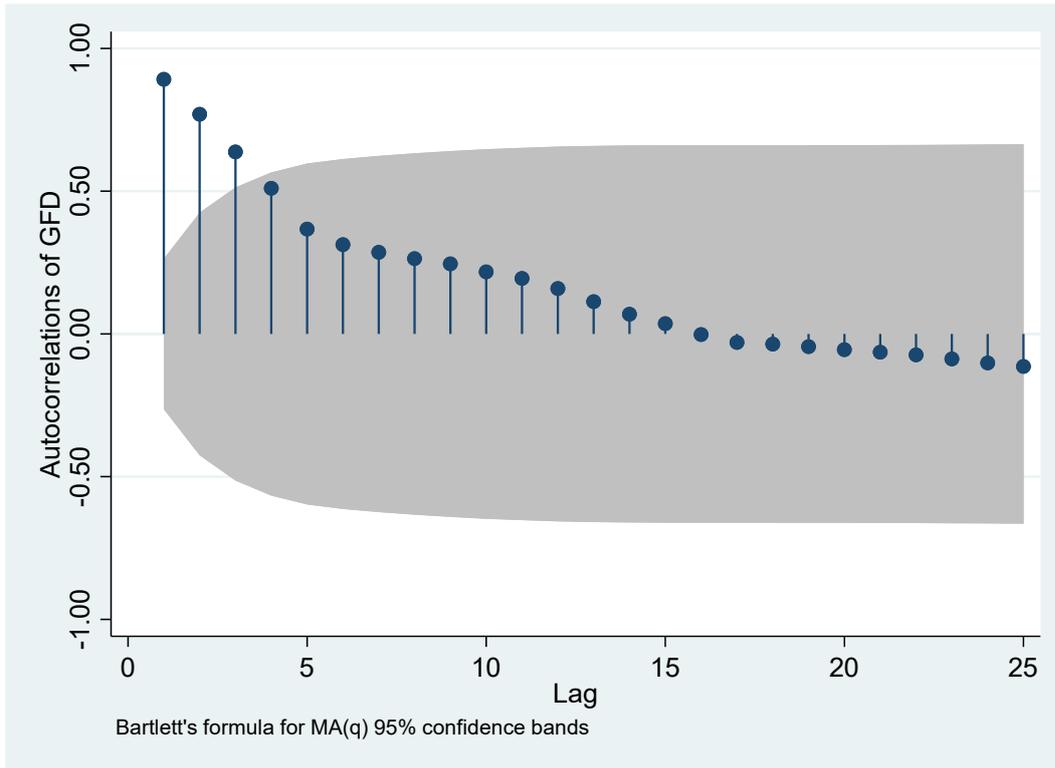

Source: Author's computation.

Table 7: Augmented Dickey-Fuller test for unit root

| Variables | Test Statistics Z (t) | P-Value | 1 % Critical Value | 5 % Critical Value | 10 % Critical Value |
|---|---|---|---|---|---|
| $Gross\ Fiscal\ Deficit$ | 0.702 | 0.9899 | -3.574 | -2.927 | -2.598 |
| $Gross\ Fiscal\ Deficit_{t-1}$ | -7.231 | 0.0000*** | -3.576 | -2.928 | -2.599 |

Note: *p<0.01, **p<0.05, ***p < 0.001

Source: Author's computation.

Table 8: Phillips-Perron test for unit root

| Variables | Test Statistics | | P-Value | 1 % Critical Value | 5 % Critical Value | 10 % Critical Value |
|---|---|---|---|---|---|---|
| $Gross\ Fiscal\ Deficit$ | Z(rho) | 1.670 | 0.9913 | -18.972 | -13.332 | -10.724 |
| | Z(t) | 0.782 | | -3.574 | -2.927 | -2.598 |
| $Gross\ Fiscal\ Deficit_{t-1}$ | Z(rho) | -57.386*** | 0.0000 | -18.954 | -13.324 | -10.718 |
| | Z(t) | -7.249*** | | -3.576 | -2.928 | -2.599 |

Note: *p<0.01, **p<0.05, ***p < 0.001

Source: Author's computation.

Figure 10: First Difference Gross Fiscal Deficit

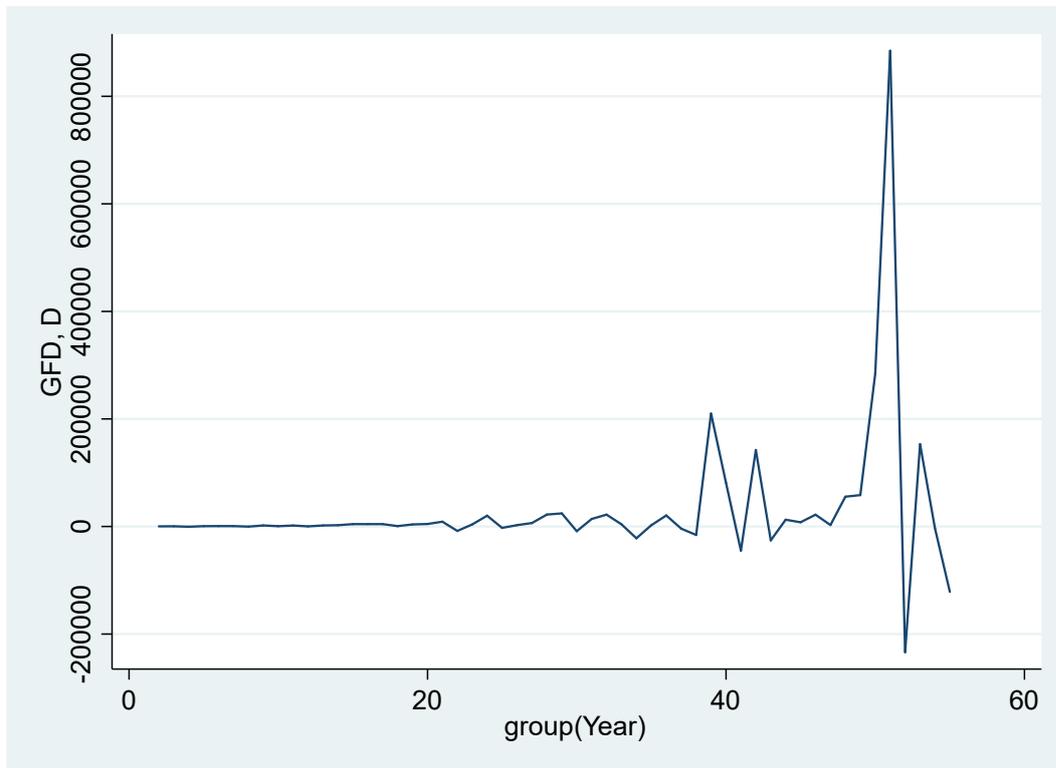

Source: Author's computation.

Table 9: ARIMA models with their AIC and BIC criteria for the entire period (1971-2025)

| p | d | q | AIC | BIC |
|---|---|---|---|---|
| 0 | 0 | 0 | 1767.478 | 1771.492 |
| 0 | 0 | 1 | 1566.679 | 1572.701 |
| 0 | 0 | 2 | 1554.43 | 1562.459 |
| 0 | 0 | 3 | 1548.794 | 1558.831 |
| 0 | 0 | 4 | 1533.113 | 1545.157 |
| 0 | 1 | 0 | 1434.093 | 1436.082 |
| 0 | 1 | 1 | 1436.206 | 1440.184 |
| 0 | 1 | 2 | 1437.779 | 1443.746 |
| 0 | 1 | 3 | 1439.636 | 1447.592 |
| 0 | 1 | 4 | 1442.22 | 1452.165 |
| 1 | 0 | 0 | 1467.834 | 1473.856 |
| 1 | 0 | 1 | 1469.824 | 1477.853 |
| 1 | 0 | 2 | 1471.169 | 1481.206 |

| 1 | 0 | 3 | 1472.466 | 1484.51 |
| 1 | 0 | 4 | 1474.809 | 1488.86 |
| 1 | 1 | 0 | 1436.049 | 1440.027 |
| 1 | 1 | 1 | 1437.199 | 1443.166 |
| 1 | 1 | 2 | 1439.405 | 1447.361 |
| 1 | 1 | 3 | 1440.663 | 1450.608 |
| 1 | 1 | 4 | 1442.124 | 1454.058 |
| 2 | 0 | 0 | 1469.728 | 1477.758 |
| 2 | 0 | 1 | 1471.423 | 1481.459 |
| 2 | 0 | 2 | 1472.374 | 1484.418 |
| 2 | 0 | 3 | 1473.954 | 1488.005 |
| 2 | 0 | 4 | 1476.42 | 1492.479 |
| 2 | 1 | 0 | 1437.342 | 1443.309 |
| 2 | 1 | 1 | 1439.127 | 1447.083 |
| 2 | 1 | 2 | 1439.837 | 1449.782 |
| 2 | 1 | 3 | 1442.513 | 1454.447 |
| 2 | 1 | 4 | 1442.965 | 1456.888 |
| 3 | 0 | 0 | 1470.731 | 1480.768 |
| 3 | 0 | 1 | 1473.42 | 1485.464 |
| 3 | 0 | 2 | 1473.05 | 1487.101 |
| 3 | 0 | 3 | 1474.377 | 1490.436 |
| 3 | 0 | 4 | 1473.434 | 1491.5 |
| 3 | 1 | 0 | 1438.46 | 1446.416 |
| 3 | 1 | 1 | 1438.734 | 1448.679 |
| 3 | 1 | 2 | 1440.843 | 1452.777 |
| 3 | 1 | 3 | 1444.241 | 1458.164 |
| 3 | 1 | 4 | 1443.308 | 1459.22 |
| 4 | 0 | 0 | 1471.416 | 1483.46 |
| 4 | 0 | 1 | 1472.296 | 1486.348 |
| 4 | 0 | 2 | 1473.317 | 1489.376 |
| 4 | 0 | 3 | 1475.627 | 1493.693 |
| 4 | 0 | 4 | 1474.726 | 1494.8 |

| 4 | 1 | 0 | 1439.629 | 1449.574 |
|---|---|---|----------|----------|
| 4 | 1 | 1 | 1440.579 | 1452.513 |
| 4 | 1 | 2 | 1442.467 | 1456.389 |
| 4 | 1 | 3 | 1444.72  | 1460.632 |
| 4 | 1 | 4 | 1451.339 | 1469.24  |

Source: Author's computation.

The following table gives the forecast for the entire period (1971-2025) based on ARIMA (0,1,0) along with the figure.

Table 10: Forecast based on the entire period (1971-2025)

| Year | Forecast |
|------|----------|
| 2026 | 1643162 |
| 2027 | 1673012 |
| 2028 | 1702862 |
| 2029 | 1732712 |
| 2030 | 1762562 |
| 2031 | 1792412 |
| 2032 | 1822263 |
| 2033 | 1852113 |
| 2034 | 1881963 |
| 2035 | 1911813 |
| 2036 | 1941663 |
| 2037 | 1971513 |
| 2038 | 2001363 |
| 2039 | 2031213 |
| 2040 | 2061063 |
| 2041 | 2090913 |
| 2042 | 2120763 |
| 2043 | 2150613 |
| 2044 | 2180463 |
| 2045 | 2210313 |
| 2046 | 2240164 |

| 2047 | 2270014 |

Source: Author's computation.

Based on the forecasted values, the Gross Fiscal Deficit is expected to increase by an average of 1% annually. This trend remains consistent for the forecast based on the sub-period.

Figure 11: Gross Fiscal Deficit forecast based on the entire period (1971-2025)

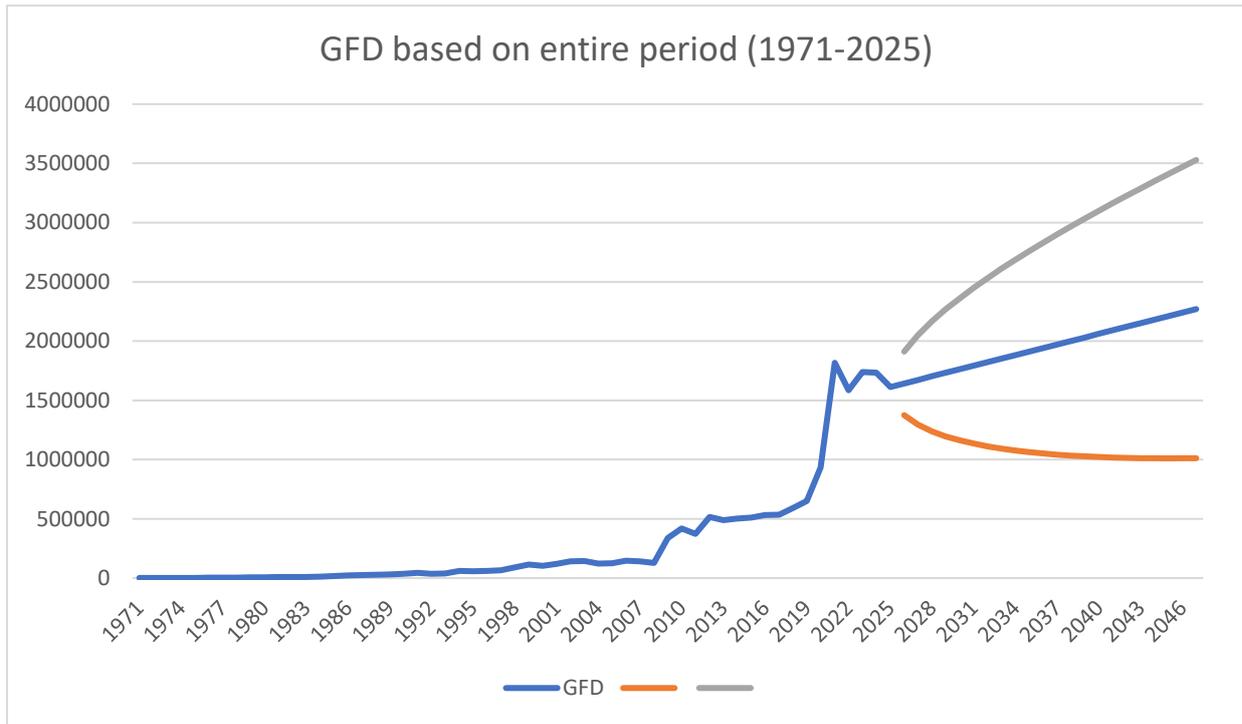

Source: Author's computation.

Table 11: ARIMA models with their AIC and BIC criteria for the sub-period (1991-2025)

| p | d | q | AIC | BIC |
|---|---|---|---|---|
| 0 | 0 | 0 | 1118.414 | 1121.525 |
| 0 | 0 | 1 | 1007.375 | 1012.041 |
| 0 | 0 | 2 | 1000.718 | 1006.94 |
| 0 | 0 | 3 | 994.8445 | 1002.621 |
| 0 | 0 | 4 | 980.5018 | 989.8339 |
| 0 | 1 | 0 | 919.4196 | 920.9459 |
| 0 | 1 | 1 | 921.6391 | 924.6918 |
| 0 | 1 | 2 | 923.915 | 928.4941 |
| 0 | 1 | 3 | 926.9552 | 933.0606 |
| 0 | 1 | 4 | 931.1374 | 938.7692 |
| 1 | 0 | 0 | 952.9099 | 957.576 |
| 1 | 0 | 1 | 954.9958 | 961.2172 |
| 1 | 0 | 2 | 957.0111 | 964.7878 |
| 1 | 0 | 3 | 959.3688 | 968.7009 |
| 1 | 0 | 4 | 963.409 | 974.2964 |
| 1 | 1 | 0 | 921.3931 | 924.4459 |
| 1 | 1 | 1 | 923.0046 | 927.5837 |
| 1 | 1 | 2 | 925.7031 | 931.8086 |
| 1 | 1 | 3 | 928.5155 | 936.1473 |
| 1 | 1 | 4 | 931.2948 | 940.4529 |
| 2 | 0 | 0 | 954.8262 | 961.0476 |
| 2 | 0 | 1 | 956.7714 | 964.5481 |
| 2 | 0 | 2 | 958.3075 | 967.6396 |
| 2 | 0 | 3 | 961.3671 | 972.2545 |
| 2 | 0 | 4 | 964.9622 | 977.4049 |
| 2 | 1 | 0 | 922.9813 | 927.5604 |
| 2 | 1 | 1 | 924.947 | 931.0525 |
| 2 | 1 | 2 | 926.9482 | 934.58 |
| 2 | 1 | 3 | 930.4966 | 939.6547 |
| 2 | 1 | 4 | 931.3332 | 942.0177 |
| 3 | 0 | 0 | 956.1026 | 963.8793 |
| 3 | 0 | 1 | 957.7178 | 967.0499 |
| 3 | 0 | 2 | 960.3186 | 971.206 |
| 3 | 0 | 3 | 963.1725 | 975.6153 |
| 3 | 0 | 4 | 963.2799 | 977.278 |
| 3 | 1 | 0 | 924.4807 | 930.5861 |
| 3 | 1 | 1 | 925.7341 | 933.3659 |
| 3 | 1 | 2 | 928.5465 | 937.7046 |
| 3 | 1 | 3 | 932.3076 | 942.9921 |
| 3 | 1 | 4 | 933.2907 | 945.5016 |
| 4 | 0 | 0 | 957.1499 | 966.482 |

| 4 | 0 | 1 | 958.9204 | 969.8078 |
| 4 | 0 | 2 | 961.3303 | 973.7731 |
| 4 | 0 | 3 | 964.7975 | 978.7956 |
| 4 | 0 | 4 | 964.7313 | 980.2848 |
| 4 | 1 | 0 | 926.0456 | 933.6774 |
| 4 | 1 | 1 | 927.7838 | 936.9419 |
| 4 | 1 | 2 | 930.403 | 941.0875 |
| 4 | 1 | 3 | 933.6269 | 945.8378 |
| 4 | 1 | 4 | 933.5717 | 947.3089 |

Source: Author's computation.

Table 12: Forecast based on the sub-period (1991-2025)

| Year | Forecast |
| --- | --- |
| 2026 | 1643162 |
| 2027 | 1673012 |
| 2028 | 1702862 |
| 2029 | 1732712 |
| 2030 | 1762562 |
| 2031 | 1792412 |
| 2032 | 1822263 |
| 2033 | 1852113 |
| 2034 | 1881963 |
| 2035 | 1911813 |
| 2036 | 1941663 |
| 2037 | 1971513 |
| 2038 | 2001363 |
| 2039 | 2031213 |
| 2040 | 2061063 |
| 2041 | 2090913 |
| 2042 | 2120763 |
| 2043 | 2150613 |
| 2044 | 2180463 |
| 2045 | 2210313 |

| 2046 | 2240164 |
| 2047 | 2270014 |

Source: Author's computation.

Figure 12: Gross Fiscal Deficit forecast based on the sub period (1991-2025)

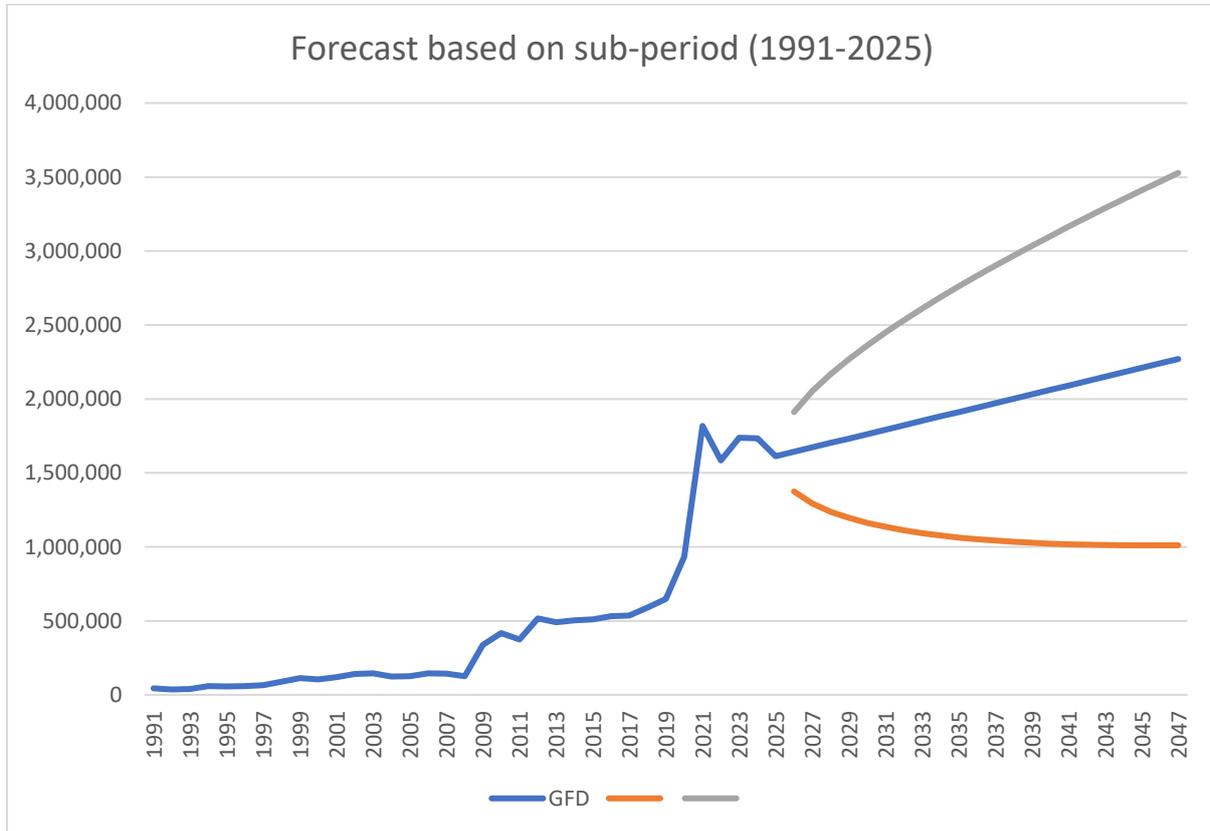

Source: Author's computation.

There is no difference in the forecasted values between the entire period and the sub-period, nor in the ARIMA models selected based on AIC and BIC criteria.

All the above steps are repeated for Gross Domestic Product (GDP). Variable GDP is extracted from the RBI database in Rs. Crores term.

Figure 13: Gross Domestic Product (in Rs. Crores)

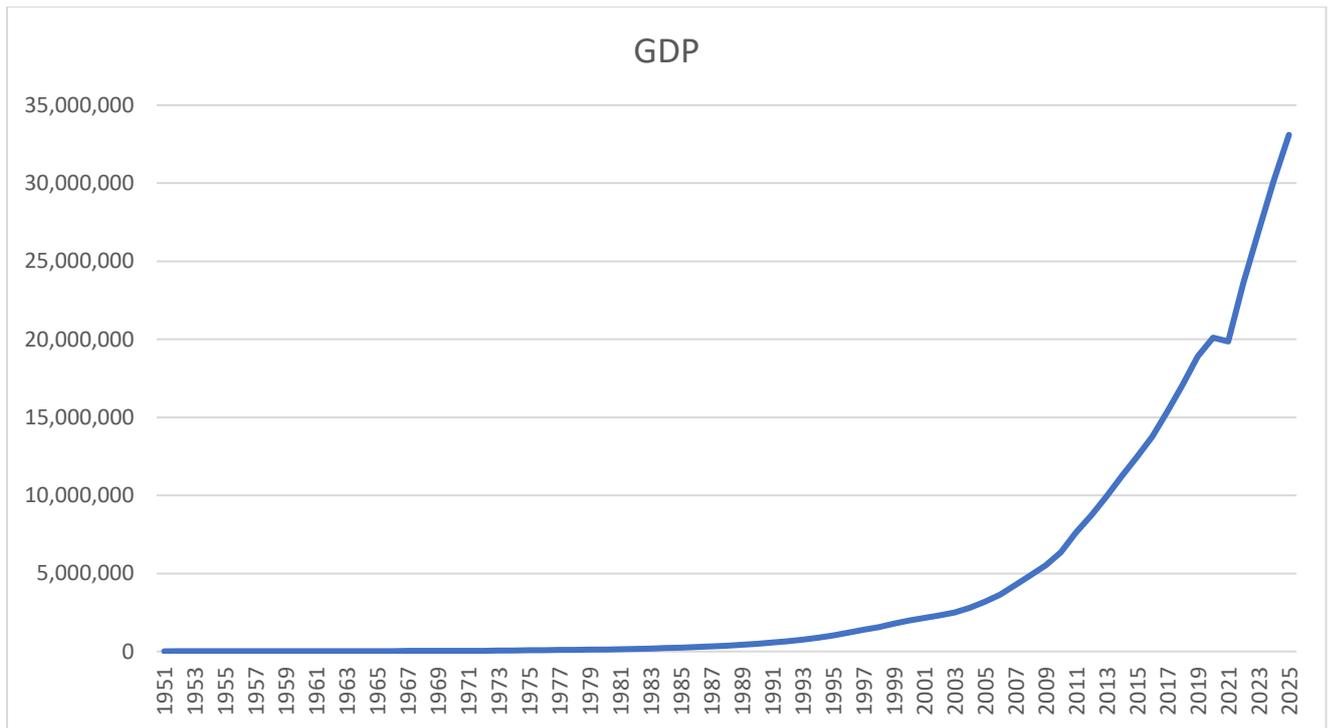

Source: RBI Database.

Figure 14: Partial Autocorrelation Function (PACF) for level form GDP

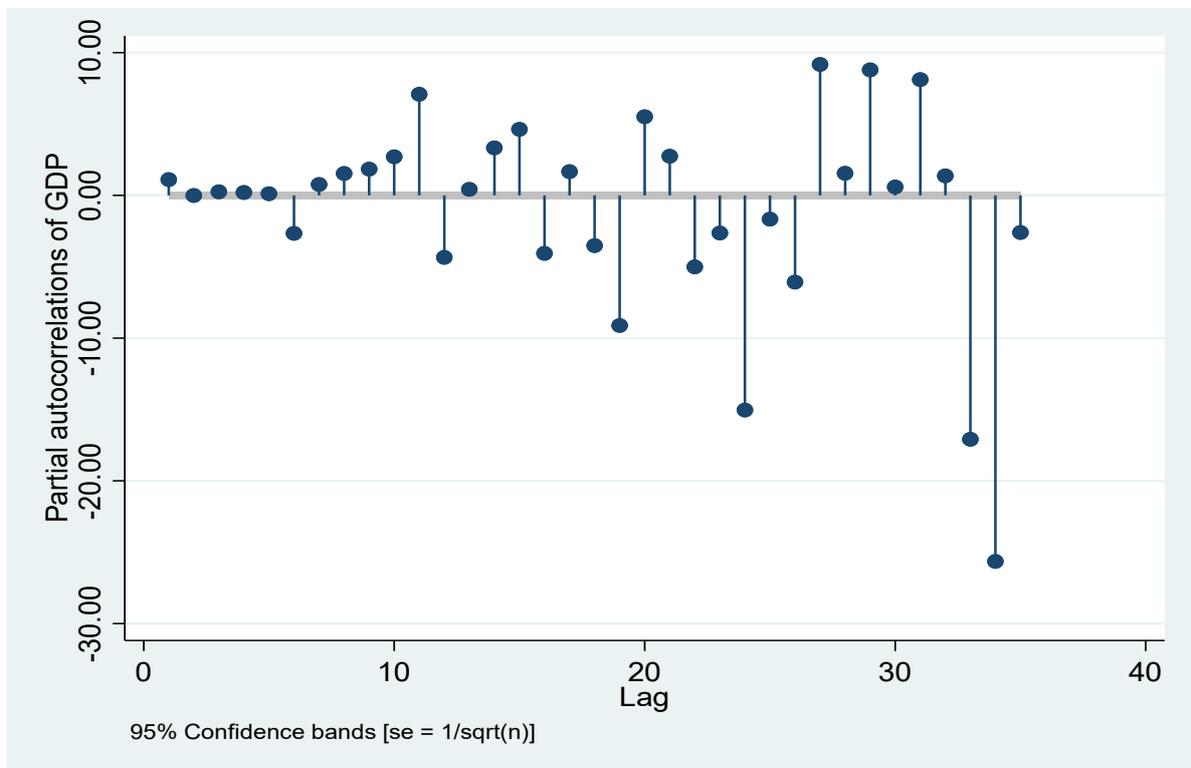

Source: Author's computation.

Figure 15: Autocorrelation Function (ACF) for level form GDP

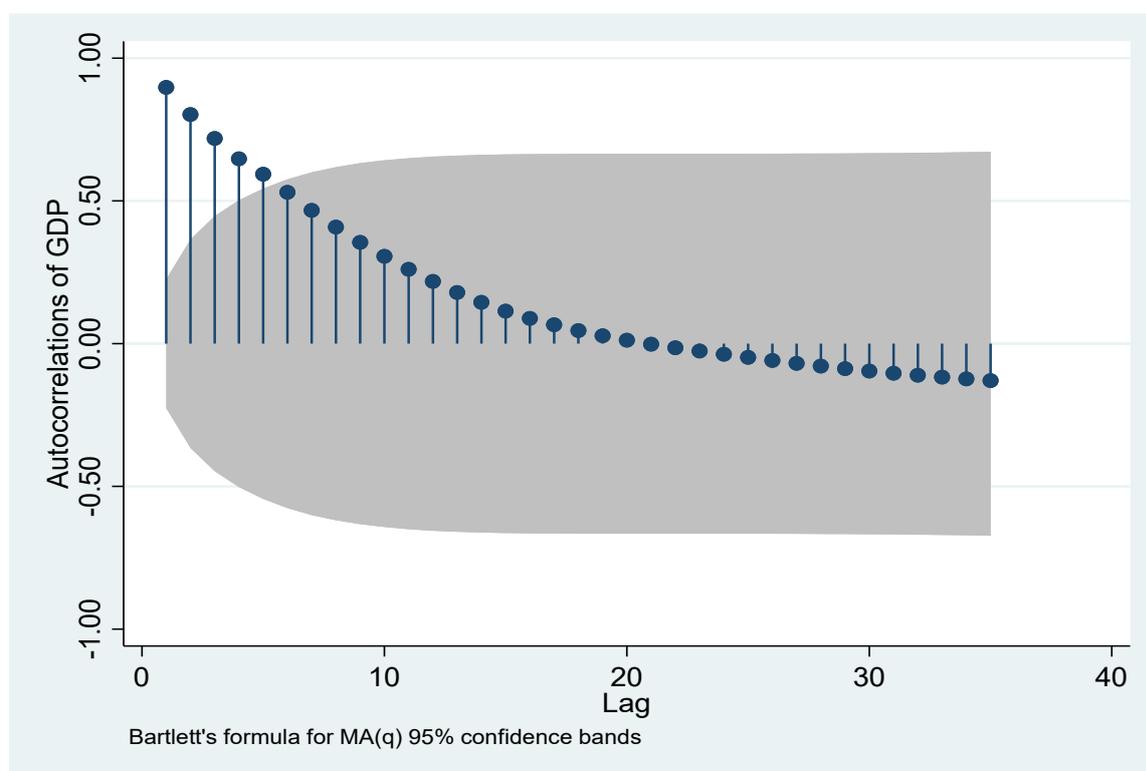

Source: Author's computation.

Table 13: Augmented Dicky Fuller test for unit root

| Variables | Test Statistics Z (t) | P-Value | 1 % Critical Value | 5 % Critical Value | 10 % Critical Value |
|---|---|---|---|---|---|
| *Gross Domestic Product* | 16.714 | 1.0000 | -3.546 | -2.911 | -2.59 |
| *Gross Domestic Product$_{t-1}$* | -1.878 | 0.3425 | -3.548 | -2.912 | -2.591 |
| *Gross Domestic Product$_{t-2}$* | -12.141 | 0.0000*** | -3.549 | -2.912 | -2.591 |

Note: *p<0.01, **p<0.05, ***p < 0.001

Source: Author's computation.

Table 14: Phillips Perron test for unit root

| Variables | Test Statistics | | P-Value | 1 % Critical Value | 5 % Critical Value | 10 % Critical Value |
|---|---|---|---|---|---|---|
| *Gross Domestic Product* | Z(rho) | 7.902 | 1.0000 | -19.332 | -13.492 | -10.84 |
| | Z(t) | 19.974 | | -3.546 | -2.911 | -2.59 |
| *Gross Domestic Product$_{t-1}$* | Z(rho) | -5.372 | 0.6803 | -19.314 | -13.484 | -10.83 |
| | Z(t) | -1.184 | | -3.548 | -2.912 | -2.59 |
| *Gross Domestic Product$_{t-2}$* | Z(rho) | -82.192 | 0.00000*** | -19.296 | -13.476 | -10.832 |
| | Z(t) | -14.315 | | -3.549 | -2.912 | -2.591 |

Note: *p<0.01, **p<0.05, ***p < 0.001

Source: Author's computation.

Figure 16: First Difference GDP

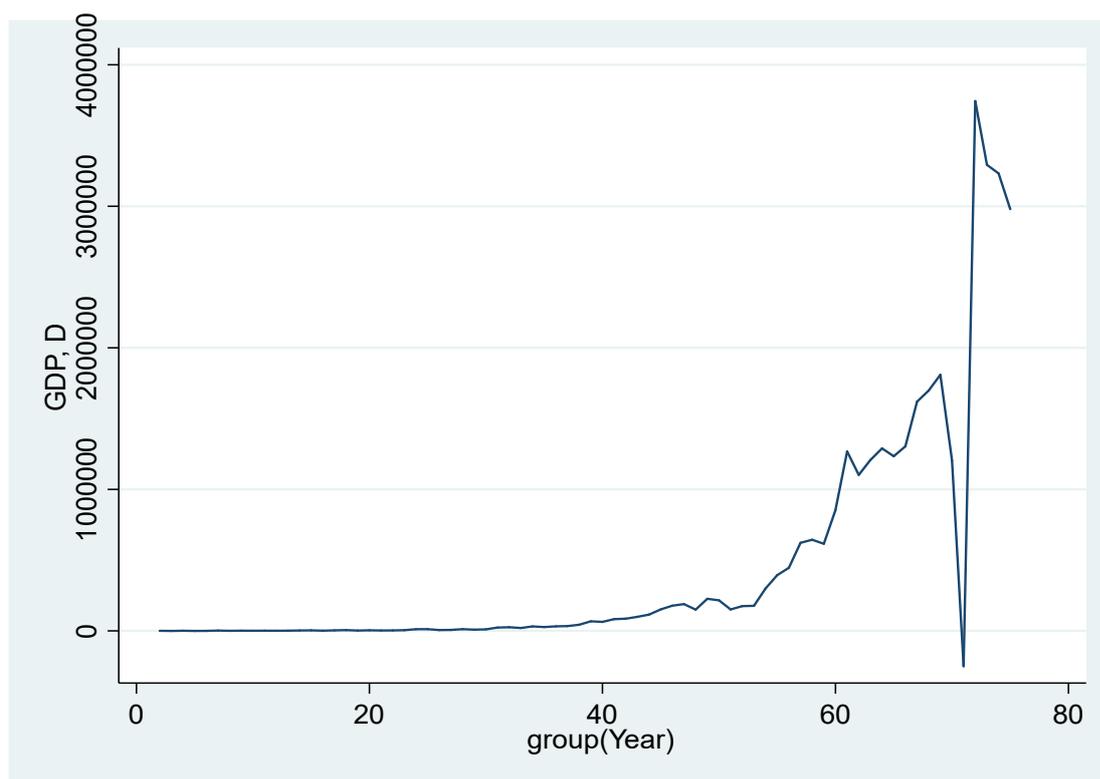

Source: Author's computation.

Figure 17: Second Difference GDP

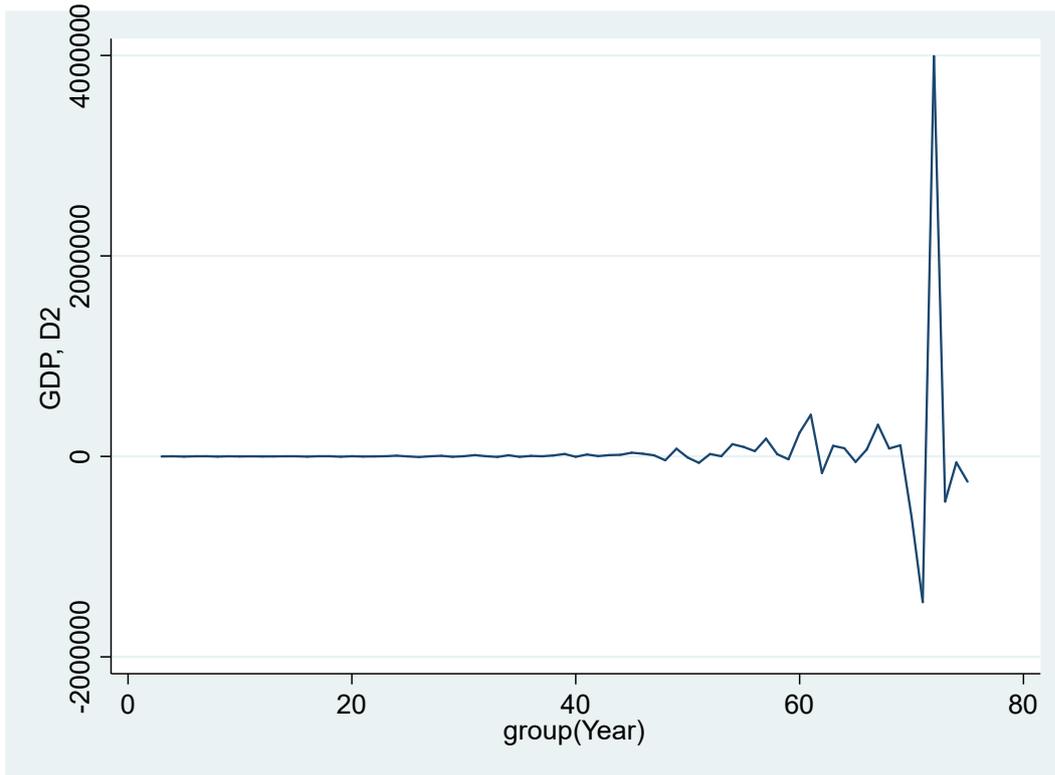

Source: Author's computation.

Table 15: ARIMA models with their AIC and BIC criteria based on the entire period (1951-2025)

| p | d | q | AIC | BIC |
|---|---|---|---|---|
| 0 | 0 | 0 | 2845.459 | 2850.094 |
| 0 | 0 | 1 | 2547.868 | 2554.82 |
| 0 | 0 | 2 | 2537.67 | 2546.94 |
| 0 | 1 | 0 | 2248.632 | 2250.936 |
| 0 | 1 | 1 | 2214.287 | 2218.895 |
| 0 | 1 | 2 | 2215.622 | 2222.535 |
| 0 | 2 | 0 | 2128.639 | 2130.93 |
| 0 | 2 | 1 | 2114.332 | 2118.913 |
| 0 | 2 | 2 | 2116.974 | 2123.845 |
| 1 | 0 | 0 | 2325.949 | 2332.901 |
| 1 | 0 | 1 | 2249.034 | 2258.304 |
| 1 | 0 | 2 | 2233.088 | 2244.676 |
| 1 | 1 | 0 | 2156.946 | 2161.554 |
| 1 | 1 | 1 | 2144.673 | 2151.585 |
| 1 | 1 | 2 | 2147.254 | 2156.471 |
| 1 | 2 | 0 | 2121.198 | 2125.779 |
| 1 | 2 | 1 | 2115.911 | 2122.783 |
| 1 | 2 | 2 | 2118.682 | 2127.844 |

Source: Author's computation.

Table 16: ARIMA models with their AIC and BIC criteria based on sub-period (1991-2025)

| p | d | q | AIC | BIC |
|---|---|---|---|---|
| 0 | 0 | 0 | 1316.107 | 1319.218 |
| 0 | 0 | 1 | 1205.046 | 1209.712 |
| 0 | 0 | 2 | 1200.397 | 1206.618 |
| 0 | 1 | 0 | 1060.677 | 1062.204 |
| 0 | 1 | 1 | 1047.254 | 1050.306 |
| 0 | 1 | 2 | 1051.016 | 1055.595 |
| 0 | 2 | 0 | 989.6011 | 991.0976 |
| 0 | 2 | 1 | 984.4337 | 987.4267 |
| 0 | 2 | 2 | 987.7544 | 992.2439 |
| 1 | 0 | 0 | 1103.335 | 1108.001 |
| 1 | 0 | 1 | 1082.224 | 1088.446 |
| 1 | 0 | 2 | 1080.924 | 1088.701 |
| 1 | 1 | 0 | 1019.978 | 1023.03 |
| 1 | 1 | 1 | 1015.939 | 1020.518 |
| 1 | 1 | 2 | 1017.416 | 1023.522 |
| 1 | 2 | 0 | 987.3453 | 990.3383 |
| 1 | 2 | 1 | 986.2065 | 990.696 |
| 1 | 2 | 2 | 989.5833 | 995.5694 |

Source: Author's computation.

Table 17: GDP Forecast based on the entire period (1991-2025)

| Year | Forecast |
|---|---|
| 2026 | 36043867 |
| 2027 | 38984519 |

| Year | GDP |
|------|-----|
| 2028 | 41925171 |
| 2029 | 44865823 |
| 2030 | 47806475 |
| 2031 | 50747127 |
| 2032 | 53687779 |
| 2033 | 56628431 |
| 2034 | 59569083 |
| 2035 | 62509735 |
| 2036 | 65450387 |
| 2037 | 68391039 |
| 2038 | 71331691 |
| 2039 | 74272343 |
| 2040 | 77212996 |
| 2041 | 80153648 |
| 2042 | 83094300 |
| 2043 | 86034952 |
| 2044 | 88975604 |
| 2045 | 91916256 |
| 2046 | 94856908 |
| 2047 | 97797560 |

Source: Author's computation.

Figure 18: GDP Forecast based on the entire period (1951-2025)

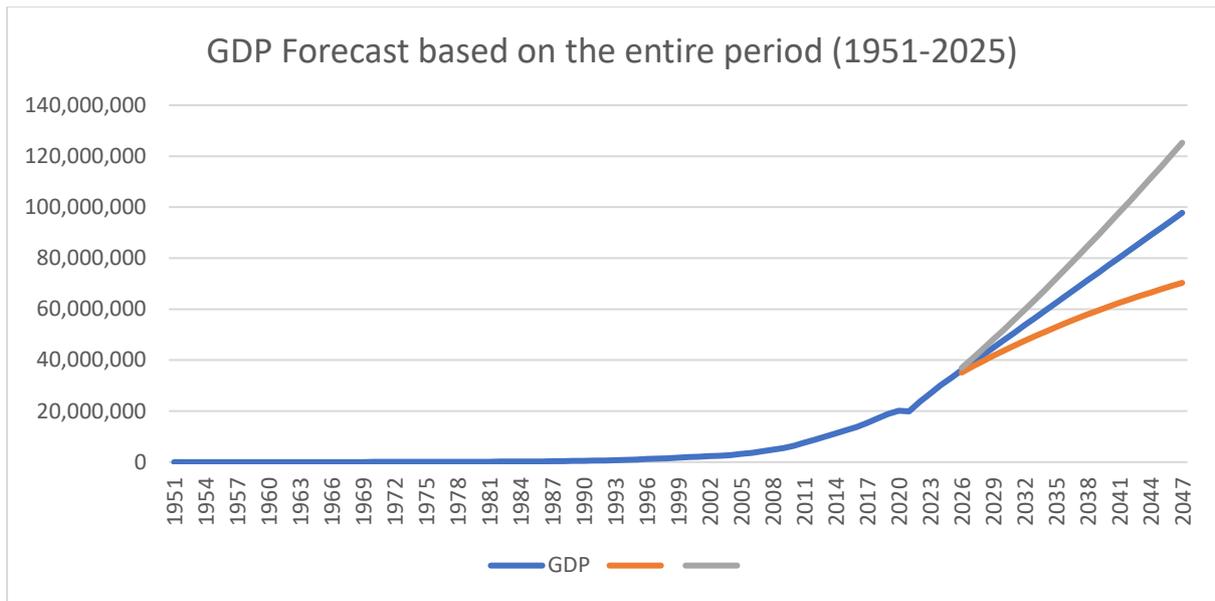

Source: Author's computation.

Table 18: Forecast based on sub-period (1991-2025)

| Year | Forecast |
|---|---|
| 2026 | 36053185 |
| 2027 | 39003155 |
| 2028 | 41953126 |
| 2029 | 44903096 |
| 2030 | 47853067 |
| 2031 | 50803037 |
| 2032 | 53753008 |
| 2033 | 56702978 |
| 2034 | 59652948 |
| 2035 | 62602919 |
| 2036 | 65552889 |
| 2037 | 68502860 |
| 2038 | 71452830 |
| 2039 | 74402800 |
| 2040 | 77352771 |
| 2041 | 80302741 |
| 2042 | 83252712 |

| 2043 | 86202682 |
| 2044 | 89152653 |
| 2045 | 92102623 |
| 2046 | 95052593 |
| 2047 | 98002564 |

Source: Author's computation.

Figure 19: Forecast based on sub-period (1991-2025)

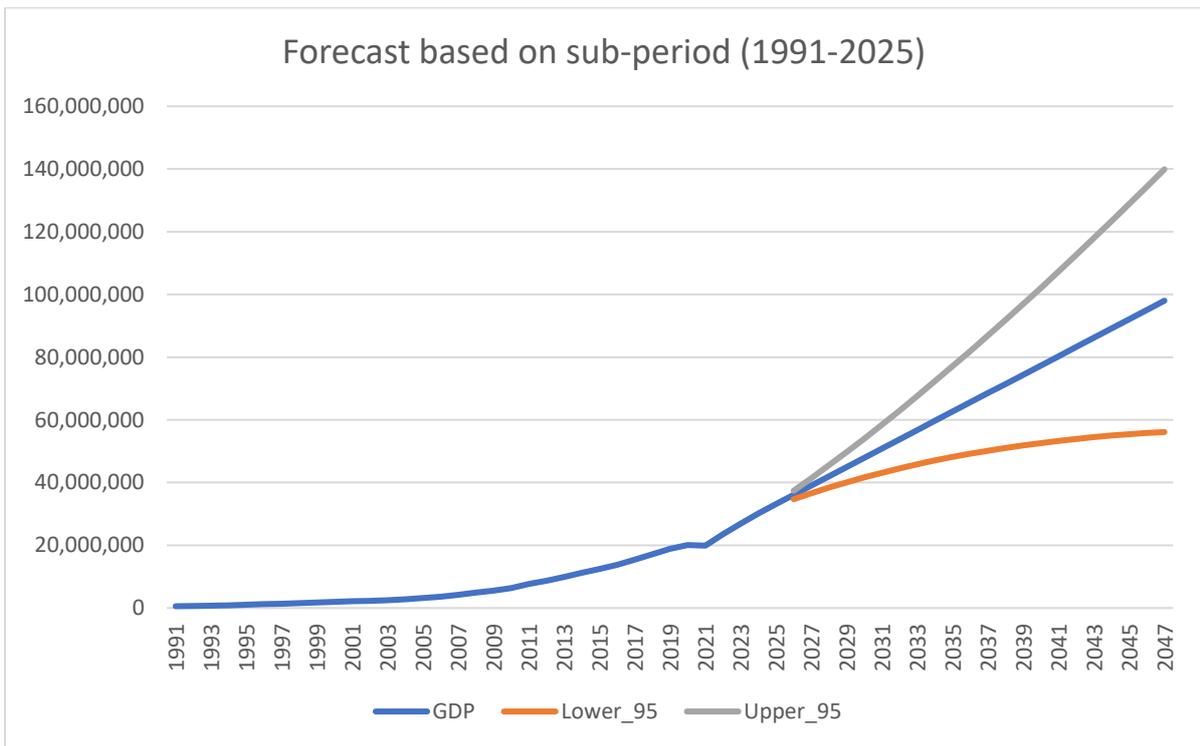

Source: Author's computation.

Based on the forecasted values, the Gross Domestic Product is expected to increase by an average of 5% annually. This trend remains consistent for the forecast based on the sub-period.

Figure 20: Gross Fiscal Deficit as % of GDP

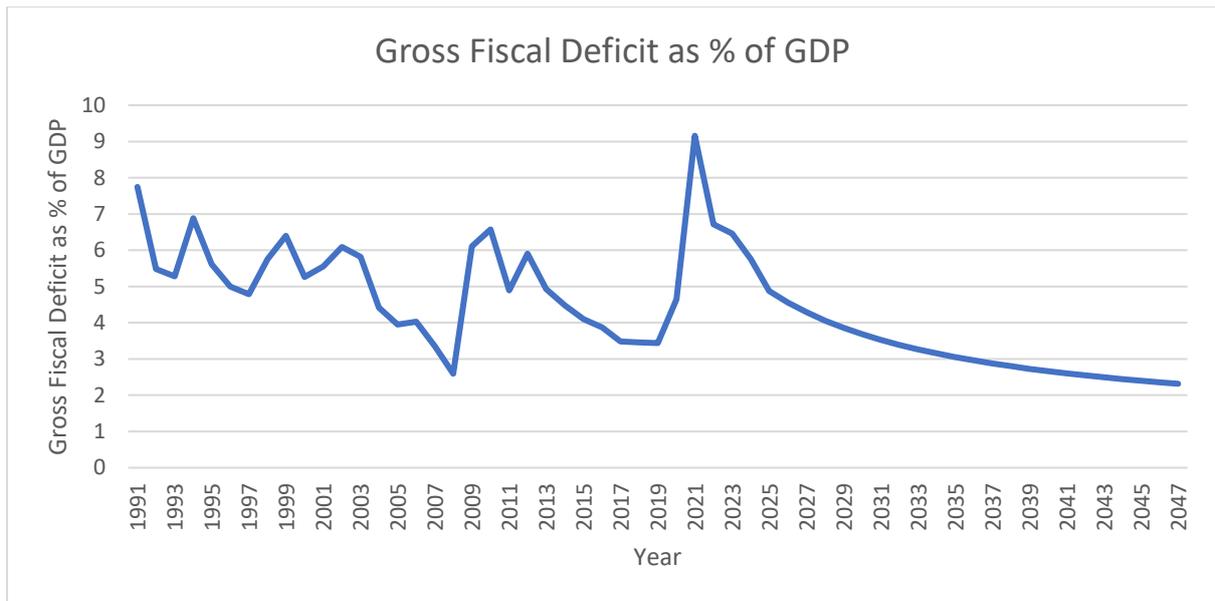

Source: Author's computation based on RBI database.

Based on data from 1991-2025, Gross Fiscal Deficit as % of GDP will decline. In 2047, it will be 2.32% of GDP.

GDP in $ terms can be found out by multiplying GDP in Indian Rupee terms by the respective exchange rate. Annual exchange rate data is taken from the RBI website. Data was available from 1971 to 2024. Of that data, the Annual Average of US $ is considered.

Figure 21: Exchange Rate

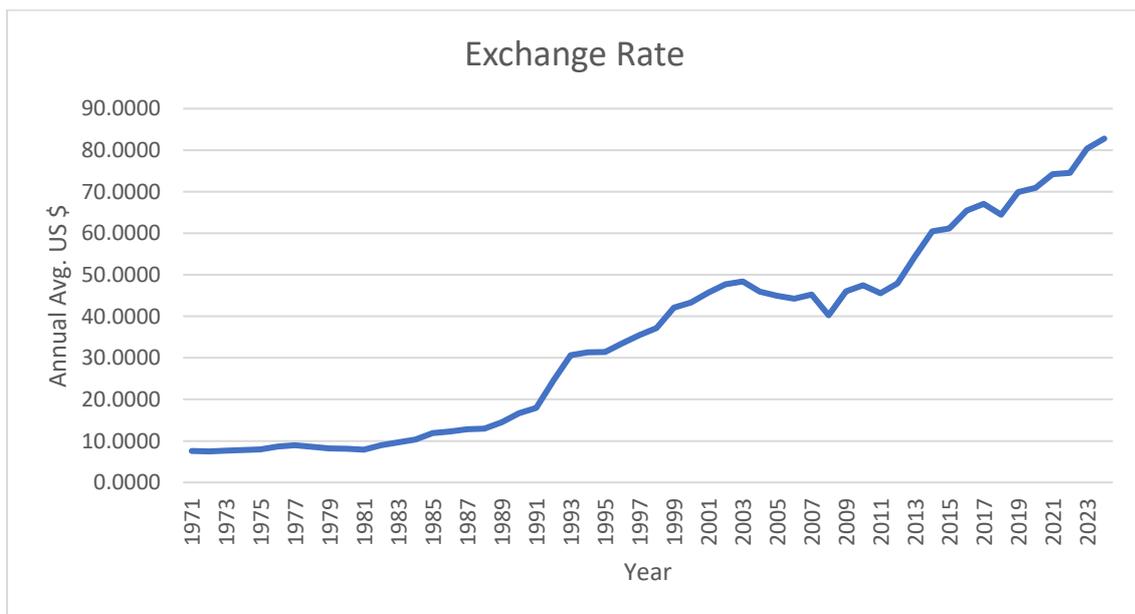

Source: RBI Website.

As is seen from the figure exchange rate is not stationary, as it is showing a positive trend.

To identify Autoregressive and moving average components Partial Autocorrelation Function (PACF) and Autocorrelation Function (ACF) are constructed. They are as follows

Figure 22: Partial Autocorrelation Function (PACF) of level form exchange rate.

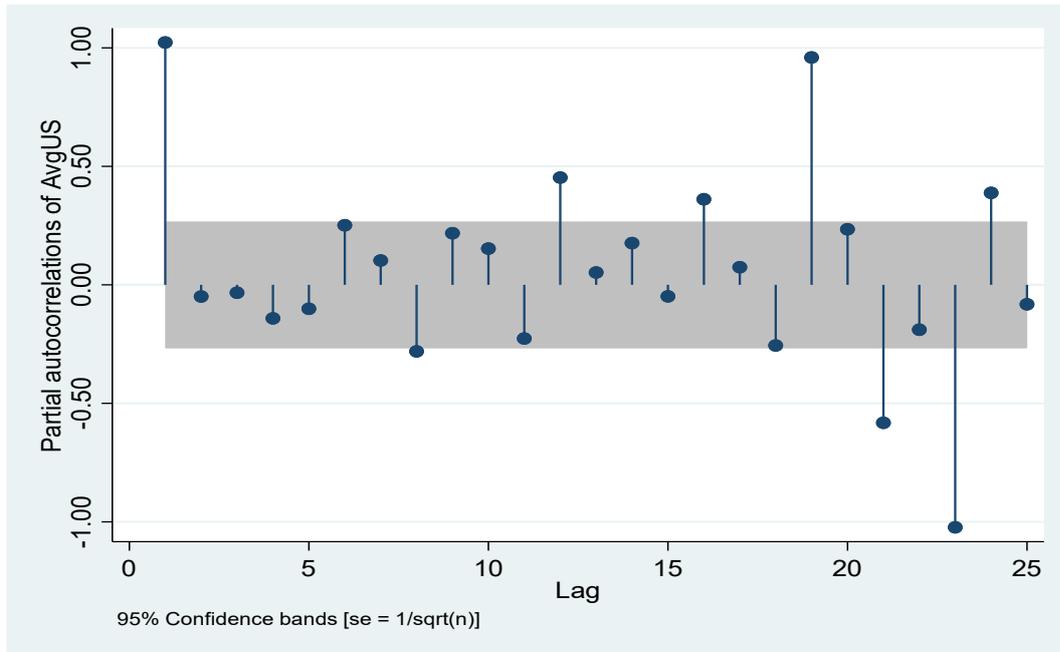

Source: Author's computation.

Figure 23: Autocorrelation Function (ACF) of level form exchange rate.

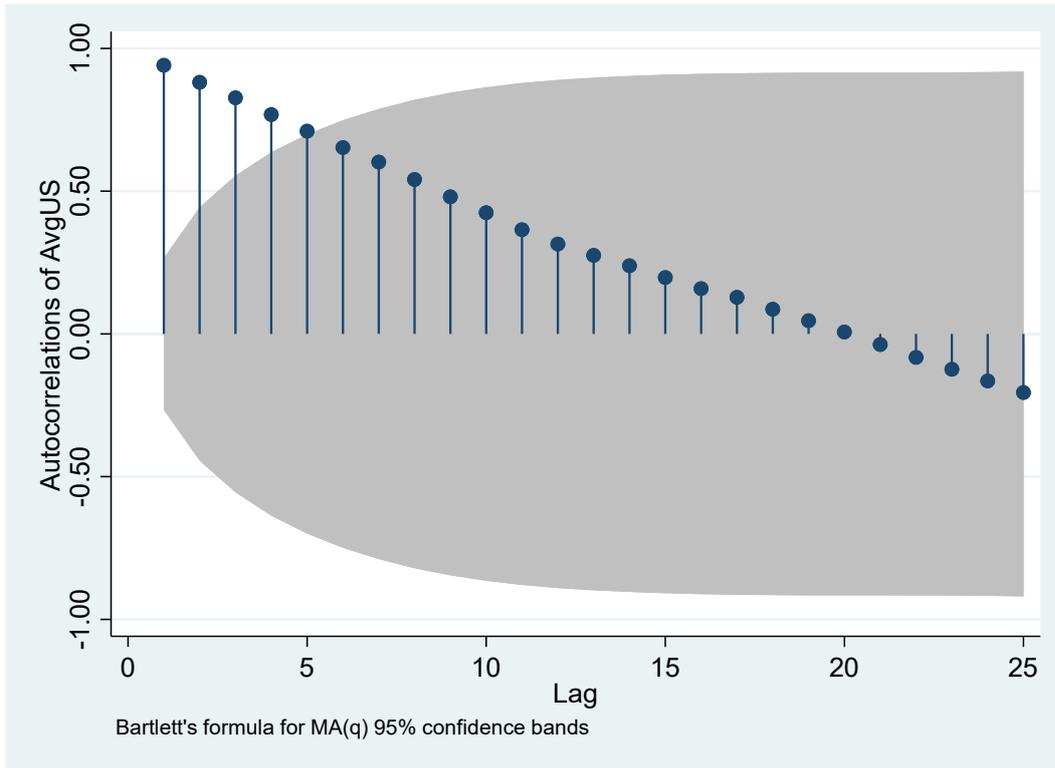

Source: Author's computation.

PACF is showing one spike. ACF is showing gradual decay after 1st lag. Further, stationarity is found by performing Augmented Dicky Fuller (DF) and Phillips Perron (PP) tests.

Table 19: Augmented Dicky Fuller unit root test.

| Variables | Test Statistics Z (t) | P-Value | 1 % Critical Value | 5 % Critical Value | 10 % Critical Value |
|---|---|---|---|---|---|
| $Exchange\ Rate$ | 1.567 | 0.9978 | -3.576 | -2.928 | -2.599 |
| $Exchange\ Rate_{t-1}$ | -6.363*** | 0.0000 | -3.577 | -2.928 | -2.599 |

Note: *p<0.01, **p<0.05, ***p < 0.001

Source: Author's computation.

Results of the Augmented Dicky Fuller unit root test showed that for the level form exchange rate, test statistics i.e. Z(t) lie beyond the confidence interval, and the P-value is also greater than 0.05. Hence, we failed to reject the null hypothesis (time series data is non-stationary). Further, the test is performed on the first difference exchange rate and the result showed that the differenced Per Capita GNI is stationary at I (1) as test statists lie in the confidence interval and the P-value is less than 0.05. Hence, we are rejecting the null of time series data as non-stationary.

Table 20: Phillips Parron test for unit root.

| Variables | Test Statistics | | P-Value | 1 % Critical Value | 5 % Critical Value | 10 % Critical Value |
|---|---|---|---|---|---|---|
| Exchange Rate | Z(rho) | 1.159 | 0.9971 | -18.954 | -13.324 | -10.718 |
| | Z(t) | 1.393 | | -3.576 | -2.928 | -2.599 |
| Exchange Rate$_{t-1}$ | Z(rho) | -48.942*** | 0.0000 | -18.936 | -13.316 | -10.712 |
| | Z(t) | -6.408*** | | -3.577 | -2.928 | -2.599 |

Note: *p<0.01, **p<0.05, ***p < 0.001

Source: Author's computation.

Results of the Phillips-perron unit root test showed that the exchange rate at its level form is non-stationary as test statistics lie outside the confidence interval and the p-value is greater than 0.05. Hence, we failed to reject the null of non-stationarity. The same test is performed for difference exchange rate and found that even at 1% significance we are rejecting the null of non-stationarity and accepting the alternative of stationary time series.

In short, the results of Dicky Fuller and Phillips Perron's test indicated that the exchange rate is non-stationary at the level form but it became stationary at first difference.

Figure 24: First Difference Exchange Rate

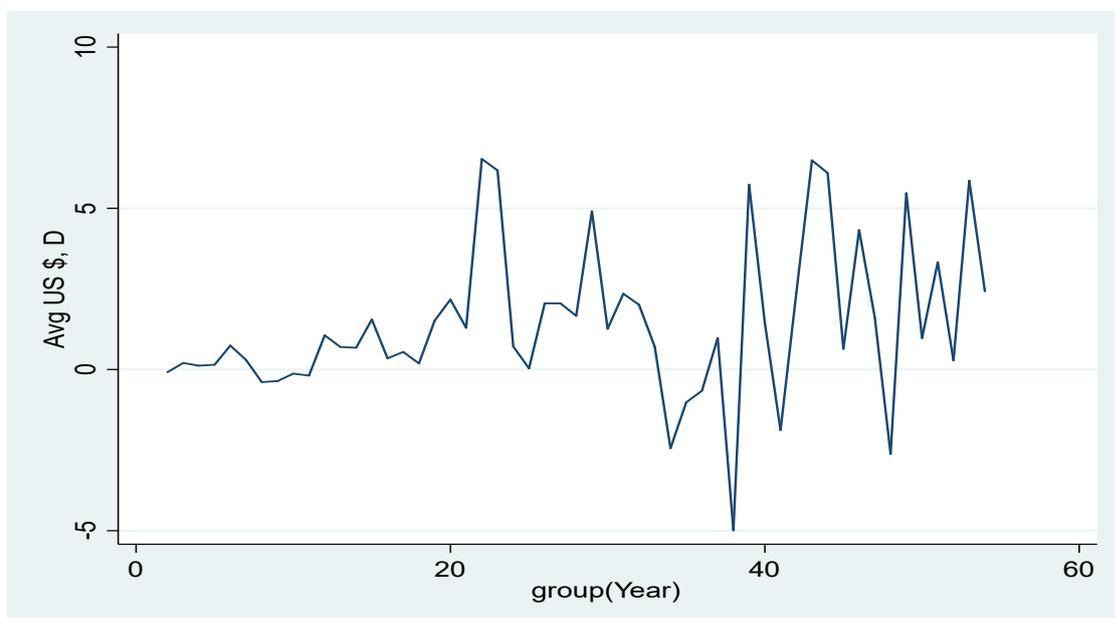

Source: Author's computation.

The exchange rate on average shows stationarity after taking the first difference. Even though there are fluctuations, there is no clear upward or downward trend. This suggests that taking the first difference is appropriate for stationarity.

For the model selection and forecasting, the entire period from which data is available, i.e., 1971-2024, was considered, and then the period after the LPG policy, i.e., 1991-2024, was considered.

ARIMA (0,1,0) model is chosen based on auto ARIMA and stepwise fit command in R and Python respectively.

Table 21: Forecast based on the entire period (1971-2024)

| Year | Forecast | Lower_95 | Upper_95 |
| --- | --- | --- | --- |
| 2025 | 84.20917 | 79.47523 | 88.94311 |
| 2026 | 85.62864 | 78.93384 | 92.32344 |
| 2027 | 87.04811 | 78.84869 | 95.24753 |
| 2028 | 88.46758 | 78.9997 | 97.93545 |
| 2029 | 89.88705 | 79.30164 | 100.4725 |
| 2030 | 91.30652 | 79.71079 | 102.9023 |
| 2031 | 92.72599 | 80.20117 | 105.2508 |
| 2032 | 94.14546 | 80.75586 | 107.5351 |
| 2033 | 95.56493 | 81.36312 | 109.7667 |
| 2034 | 96.9844 | 82.01437 | 111.9544 |
| 2035 | 98.40387 | 82.70317 | 114.1046 |
| 2036 | 99.82334 | 83.4245 | 116.2222 |
| 2037 | 101.2428 | 84.17435 | 118.3113 |
| 2038 | 102.6623 | 84.9495 | 120.3751 |
| 2039 | 104.0817 | 85.74729 | 122.4162 |
| 2040 | 105.5012 | 86.56547 | 124.437 |
| 2041 | 106.9207 | 87.40216 | 126.4392 |
| 2042 | 108.3402 | 88.25576 | 128.4246 |
| 2043 | 109.7596 | 89.12487 | 130.3944 |
| 2044 | 111.1791 | 90.00828 | 132.3499 |

| 2045 | 112.5986 | 90.90494 | 134.2922 |
| 2046 | 114.018 | 91.8139 | 136.2222 |
| 2047 | 115.4375 | 92.73434 | 138.1407 |

Source: Author's computation.

Figure 25: Forecast based on the entire period (1971-2024)

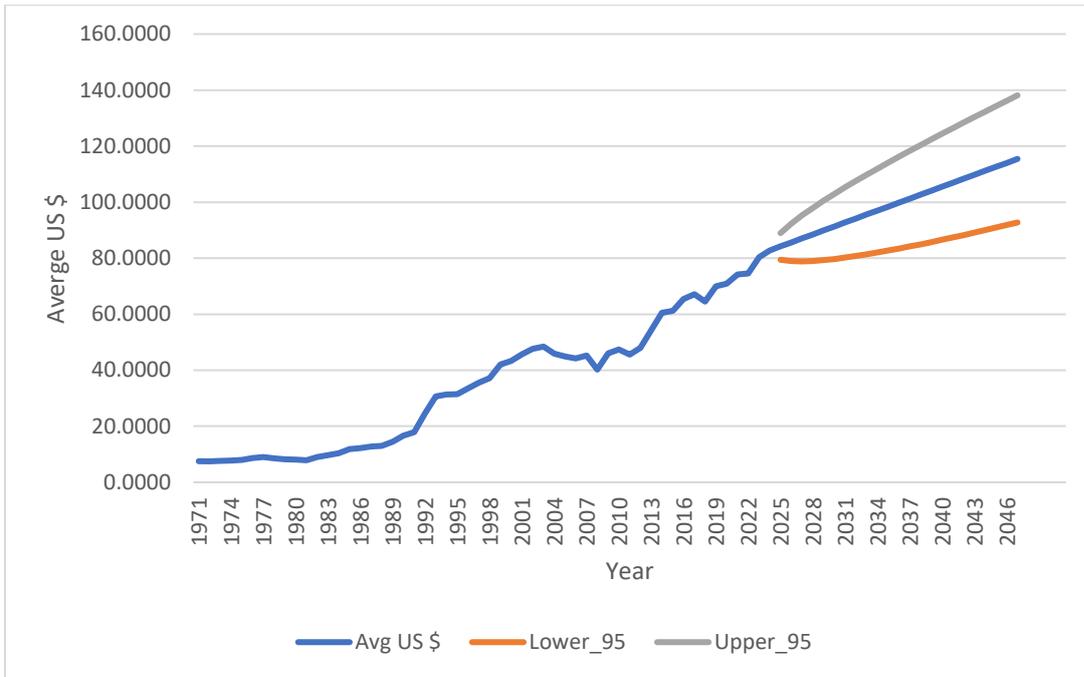

Source: Author's computation.

Table 22: Forecast based on the sub-period (1991-2024)

| Year | Forecast | Lower_95 | Upper_95 |
|---|---|---|---|
| 2025 | 84.75476 | 79.08184 | 90.42768 |
| 2026 | 86.71982 | 78.6971 | 94.74253 |
| 2027 | 88.68487 | 78.85909 | 98.51065 |
| 2028 | 90.64993 | 79.30409 | 101.9958 |
| 2029 | 92.61499 | 79.92996 | 105.3 |
| 2030 | 94.58005 | 80.68429 | 108.4758 |
| 2031 | 96.5451 | 81.53597 | 111.5542 |
| 2032 | 98.51016 | 82.46473 | 114.5556 |
| 2033 | 100.4752 | 83.45646 | 117.494 |
| 2034 | 102.4403 | 84.50093 | 120.3796 |

| 2035 | 104.4053 | 85.59039 | 123.2203 |
| 2036 | 106.3704 | 86.71883 | 126.022 |
| 2037 | 108.3354 | 87.88145 | 128.7894 |
| 2038 | 110.3005 | 89.07439 | 131.5266 |
| 2039 | 112.2656 | 90.29445 | 134.2367 |
| 2040 | 114.2306 | 91.53895 | 136.9223 |
| 2041 | 116.1957 | 92.80564 | 139.5857 |
| 2042 | 118.1607 | 94.09258 | 142.2289 |
| 2043 | 120.1258 | 95.39812 | 144.8535 |
| 2044 | 122.0909 | 96.72079 | 147.4609 |
| 2045 | 124.0559 | 98.05933 | 150.0525 |
| 2046 | 126.021 | 99.41262 | 152.6293 |
| 2047 | 127.986 | 100.7797 | 155.1924 |

Source: Author's computation.

Figure 26: Forecast based on the sub-period (1991-2024)

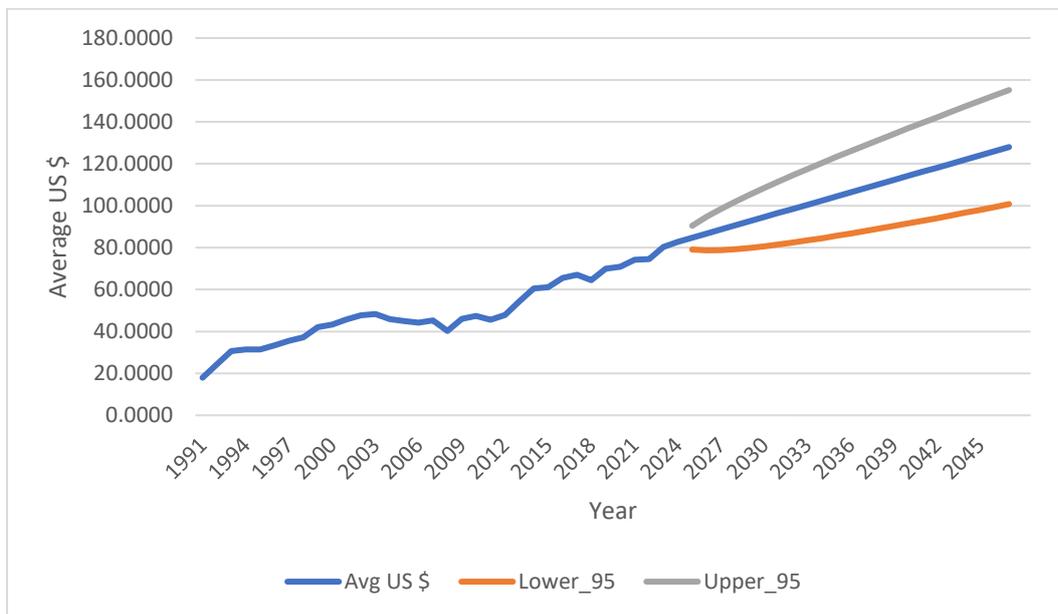

Source: Author's computation.

Based on the forecasted values, the Exchange Rate is expected to increase by an average of 1% annually based on the entire period forecast, and 2 % based on sub-period forecast.

GDP in dollar terms is found by

$$GDP\ (\$) = \frac{GDP\ (Rs. Crores)}{Exchange\ Rate}$$

Table 23: GDP ($) from 1970-2024

| Year | GDP | Exchange Rate | GDP in $ |
|---|---|---|---|
| 1971 | 46,817 | 7.5578 | 6194.5093 |
| 1972 | 50,120 | 7.4731 | 6706.7096 |
| 1973 | 55,245 | 7.6750 | 7198.0895 |
| 1974 | 67,241 | 7.7925 | 8628.8940 |
| 1975 | 79,378 | 7.9408 | 9996.2064 |
| 1976 | 85,212 | 8.6825 | 9814.2758 |
| 1977 | 91,812 | 8.9775 | 10226.8653 |
| 1978 | 1,04,024 | 8.5858 | 12115.7632 |
| 1979 | 1,12,671 | 8.2267 | 13695.8159 |
| 1980 | 1,23,562 | 8.0975 | 15259.2997 |
| 1981 | 1,47,063 | 7.9092 | 18593.8993 |
| 1982 | 1,72,776 | 8.9683 | 19265.1345 |
| 1983 | 1,93,255 | 9.6660 | 19993.2302 |
| 1984 | 2,25,074 | 10.3400 | 21767.3345 |
| 1985 | 2,52,188 | 11.8886 | 21212.6112 |
| 1986 | 2,84,534 | 12.2349 | 23255.9388 |
| 1987 | 3,18,366 | 12.7782 | 24914.7721 |
| 1988 | 3,61,865 | 12.9658 | 27909.1690 |
| 1989 | 4,29,363 | 14.4817 | 29648.6594 |
| 1990 | 4,93,278 | 16.6492 | 29627.7067 |
| 1991 | 5,76,109 | 17.9428 | 32108.0994 |
| 1992 | 6,62,260 | 24.4737 | 27060.0883 |
| 1993 | 7,61,196 | 30.6488 | 24836.0762 |
| 1994 | 8,75,992 | 31.3655 | 27928.5338 |
| 1995 | 10,27,570 | 31.3986 | 32726.6203 |
| 1996 | 12,05,583 | 33.4498 | 36041.5511 |
| 1997 | 13,94,816 | 35.4999 | 39290.7015 |

| Year | | | |
|------|------|------|------|
| 1998 | 15,45,294 | 37.1648 | 41579.5020 |
| 1999 | 17,72,297 | 42.0706 | 42126.7349 |
| 2000 | 19,88,262 | 43.3327 | 45883.6300 |
| 2001 | 21,39,886 | 45.6844 | 46840.6215 |
| 2002 | 23,15,243 | 47.6919 | 48545.8318 |
| 2003 | 24,92,614 | 48.3953 | 51505.2864 |
| 2004 | 27,92,530 | 45.9516 | 60771.1188 |
| 2005 | 31,86,332 | 44.9315 | 70915.3250 |
| 2006 | 36,32,125 | 44.2735 | 82038.3448 |
| 2007 | 42,54,629 | 45.2495 | 94025.9889 |
| 2008 | 48,98,662 | 40.2607 | 121673.5442 |
| 2009 | 55,14,152 | 45.9933 | 119890.3402 |
| 2010 | 63,66,407 | 47.4433 | 134189.7916 |
| 2011 | 76,34,472 | 45.5626 | 167560.0625 |
| 2012 | 87,36,329 | 47.9229 | 182299.6670 |
| 2013 | 99,44,013 | 54.4099 | 182761.0987 |
| 2014 | 1,12,33,522 | 60.5019 | 185672.2121 |
| 2015 | 1,24,67,959 | 61.1436 | 203912.7398 |
| 2016 | 1,37,71,874 | 65.4685 | 210358.7832 |
| 2017 | 1,53,91,669 | 67.0720 | 229479.7978 |
| 2018 | 1,70,90,042 | 64.4549 | 265147.2890 |
| 2019 | 1,88,99,668 | 69.9229 | 270292.9655 |
| 2020 | 2,01,03,593 | 70.8970 | 283560.6397 |
| 2021 | 1,98,54,096 | 74.2250 | 267485.1507 |
| 2022 | 2,35,97,399 | 74.5039 | 316727.0666 |
| 2023 | 2,68,90,473 | 80.3635 | 334610.7288 |
| 2024 | 3,01,22,956 | 82.7897 | 363849.0766 |
| 2025 | 3,31,03,215 | 84.2092 | 393107.0057 |
| 2026 | 3,60,43,867 | 85.6286 | 420932.3752 |
| 2027 | 3,89,84,519 | 87.0481 | 447850.2632 |
| 2028 | 4,19,25,171 | 88.4676 | 473904.3517 |
| 2029 | 4,48,65,823 | 89.8870 | 499135.5631 |

| 2030 | 4,78,06,475 | 91.3065 | 523582.2754 |
| 2031 | 5,07,47,127 | 92.7260 | 547280.5165 |
| 2032 | 5,36,87,779 | 94.1455 | 570264.1413 |
| 2033 | 5,66,28,431 | 95.5649 | 592564.9933 |
| 2034 | 5,95,69,083 | 96.9844 | 614213.0521 |
| 2035 | 6,25,09,735 | 98.4039 | 635236.5669 |
| 2036 | 6,54,50,387 | 99.8233 | 655662.1805 |
| 2037 | 6,83,91,039 | 101.2428 | 675515.0420 |
| 2038 | 7,13,31,691 | 102.6623 | 694818.9075 |
| 2039 | 7,42,72,343 | 104.0817 | 713596.2402 |
| 2040 | 7,72,12,996 | 105.5012 | 731868.2926 |
| 2041 | 8,01,53,648 | 106.9207 | 749655.1884 |
| 2042 | 8,30,94,300 | 108.3402 | 766975.9974 |
| 2043 | 8,60,34,952 | 109.7596 | 783848.8027 |
| 2044 | 8,89,75,604 | 111.1791 | 800290.7637 |
| 2045 | 9,19,16,256 | 112.5986 | 816318.1746 |
| 2046 | 9,48,56,908 | 114.0180 | 831946.5185 |
| 2047 | 9,77,97,560 | 115.4375 | 847190.5157 |

Source: Author's computation.

Table 24: GDP ($) from 1991-2024

| Year | GDP | Exchange Rate | GDP in $ |
| --- | --- | --- | --- |
| 1991 | 5,76,109 | 17.9428 | 32108.0994 |
| 1992 | 6,62,260 | 24.4737 | 27060.0883 |
| 1993 | 7,61,196 | 30.6488 | 24836.0762 |
| 1994 | 8,75,992 | 31.3655 | 27928.5338 |
| 1995 | 10,27,570 | 31.3986 | 32726.6203 |
| 1996 | 12,05,583 | 33.4498 | 36041.5511 |
| 1997 | 13,94,816 | 35.4999 | 39290.7015 |
| 1998 | 15,45,294 | 37.1648 | 41579.5020 |
| 1999 | 17,72,297 | 42.0706 | 42126.7349 |
| 2000 | 19,88,262 | 43.3327 | 45883.6300 |
| 2001 | 21,39,886 | 45.6844 | 46840.6215 |

| Year | | | |
|---|---|---|---|
| 2002 | 23,15,243 | 47.6919 | 48545.8318 |
| 2003 | 24,92,614 | 48.3953 | 51505.2864 |
| 2004 | 27,92,530 | 45.9516 | 60771.1188 |
| 2005 | 31,86,332 | 44.9315 | 70915.3250 |
| 2006 | 36,32,125 | 44.2735 | 82038.3448 |
| 2007 | 42,54,629 | 45.2495 | 94025.9889 |
| 2008 | 48,98,662 | 40.2607 | 121673.5442 |
| 2009 | 55,14,152 | 45.9933 | 119890.3402 |
| 2010 | 63,66,407 | 47.4433 | 134189.7916 |
| 2011 | 76,34,472 | 45.5626 | 167560.0625 |
| 2012 | 87,36,329 | 47.9229 | 182299.6670 |
| 2013 | 99,44,013 | 54.4099 | 182761.0987 |
| 2014 | 1,12,33,522 | 60.5019 | 185672.2121 |
| 2015 | 1,24,67,959 | 61.1436 | 203912.7398 |
| 2016 | 1,37,71,874 | 65.4685 | 210358.7832 |
| 2017 | 1,53,91,669 | 67.0720 | 229479.7978 |
| 2018 | 1,70,90,042 | 64.4549 | 265147.2890 |
| 2019 | 1,88,99,668 | 69.9229 | 270292.9655 |
| 2020 | 2,01,03,593 | 70.8970 | 283560.6397 |
| 2021 | 1,98,54,096 | 74.2250 | 267485.1507 |
| 2022 | 2,35,97,399 | 74.5039 | 316727.0666 |
| 2023 | 2,68,90,473 | 80.3635 | 334610.7288 |
| 2024 | 3,01,22,956 | 82.7897 | 363849.0766 |
| 2025 | 3,31,03,215 | 84.7548 | 390576.4767 |
| 2026 | 3,60,53,185 | 86.7198 | 415743.3334 |
| 2027 | 3,90,03,155 | 88.6849 | 439794.9080 |
| 2028 | 4,19,53,126 | 90.6499 | 462803.7299 |
| 2029 | 4,49,03,096 | 92.6150 | 484836.1729 |
| 2030 | 4,78,53,067 | 94.5800 | 505953.0947 |
| 2031 | 5,08,03,037 | 96.5451 | 526210.3982 |
| 2032 | 5,37,53,008 | 98.5102 | 545659.5259 |
| 2033 | 5,67,02,978 | 100.4752 | 564347.8956 |

| 2034 | 5,96,52,948 | 102.4403 | 582319.2871 |
| 2035 | 6,26,02,919 | 104.4053 | 599614.1846 |
| 2036 | 6,55,52,889 | 106.3704 | 616270.0791 |
| 2037 | 6,85,02,860 | 108.3354 | 632321.7431 |
| 2038 | 7,14,52,830 | 110.3005 | 647801.4705 |
| 2039 | 7,44,02,800 | 112.2656 | 662739.2950 |
| 2040 | 7,73,52,771 | 114.2306 | 677163.1816 |
| 2041 | 8,03,02,741 | 116.1957 | 691099.2056 |
| 2042 | 8,32,52,712 | 118.1607 | 704571.7069 |
| 2043 | 8,62,02,682 | 120.1258 | 717603.4335 |
| 2044 | 8,91,52,653 | 122.0909 | 730215.6671 |
| 2045 | 9,21,02,623 | 124.0559 | 742428.3427 |
| 2046 | 9,50,52,593 | 126.0210 | 754260.1513 |
| 2047 | 9,80,02,564 | 127.9860 | 765728.6367 |

Source: Author's computation.

**Conclusion:**

The present study tries to examine the fiscal and macroeconomic strategies essential for achieving the vision of Viksit Bharat. By taking into account constraints on domestic resource mobilization, it highlights the importance of multilateral finance institutions like World Bank, IMF, and ADB in expanding India's fiscal space through concessional financing, technical cooperation, and risk-bearing mechanisms. For the purpose of study Per Capita GNI (in Current US$) was extracted from World Bank Database. Gross Fiscal Deficit (Rs. Crores), GDP (Rs. Crores), the annual average exchange rate of the Indian Rupee to the US Dollar were extracted form RBI database. For forecasting all of these variables three step method (Identification, Estimation, and Diagnostic Checking) introduced by Box Jenkins (1970) was used. Under Identification stationarity is checked by using ACF and PACF plots. Apart from that, analysis is complemented by using ADF, and Phillips-Perron test for unit root. As all variables were non stationary at level form. So, first difference, and second difference (if required) stationarity is checked. All the variable except GDP were found stationary at first difference. ARIMA models, based on AIC and BIC criteria were used for forecasting all these variables, except in case of exchange rate where auto arima command in R and stepwise fit in python is used. Present study takes into account two periods. Starting of the first period for all variables is from where the

data is available and end period is 2023/2025. Second/Sub period is precisely from 1991 to 2023/2025.

Present study found that India's per capita GNI in 2047 (based on data 1962-2023) will become $ 5,492.2796 and for becoming a developed country it should reach above $ 14,005. As per the forecasted value of per capita GNI, the annual average growth rate is 3%. If India wants to be in the developed category status, India's per capita GNI has to grow by an annual average of 7%. India's per capita GNI in 2047 (based on data 1991-2023) will become $ 5,521.6278 and for the developed country it should reach above $ 14,005. India's Gross Fiscal Deficit (GFD) in 2047) will be Rs. 22,70,014 crores (based on 1971-2025). It remains for sub period too. As per forecasted value of GFD, the annual average growth rate is 1%. Results showed that GFD as % of GDP (based on 1991-2025) will decline and it will be 2.32% of GDP in 2047. India's GDP ($) (based on 1970-2024) in 2047 will become $ 8,47,190.5157 and for the developed country it should reach $ 21,60, 287.5375. Similarly, India's GDP ($) (based on 1991-2024) in 2047 will become 7,65,728.6367 and for the developed country, it should reach by $ 19,42,186.2439.

**References:**

Adam, C. S., & Bevan, D. L. (2005). Fiscal deficits and growth in developing countries. *Journal of Public Economics*, *89*, 571–597. https://doi.org/10.1016/j.jpubeco.2004.02.006

Ang, J. B. (2010). Does foreign aid promote growth? exploring the role of financial liberalization. *Review of Development Economics*, *14*(2), 197–212. https://doi.org/10.1111/j.1467-9361.2010.00547.x

Ansah, P. O., Campbell, E., Kotcher, J., Rosenthal, S. A., Leiserowitz, A., & Maibach, E. (2023). Predictors of U.S. public support for climate aid to developing countries. *Environmental Research Communications*, *5*(12). https://doi.org/10.1088/2515-7620/ad0ff2

Barkat, K., Alsamara, M., & Mimouni, K. (2024). Beyond economic growth goals: can foreign aid mitigate carbon dioxide emissions in developing countries? *Journal of Cleaner Production*, *471*. https://doi.org/10.1016/j.jclepro.2024.143411

Dash, A. K. (2021). Does foreign aid influence economic growth? Evidence from South Asian countries. *Transnational Corporations Review*, *15*(3) 72–85. https://doi.org/10.1080/19186444.2021.1974257

Devender, & Kumar, J. (2021). Is fiscal deficit essential for stimulating economic growth in developing economies? Theory and empirical evidence from India. *Indian Journal of Economics and Development*, *17*(3), 598–607. https://doi.org/10.35716/IJED/20294

Dreher, A., Nunnenkamp, P., & Thiele, R. (2008). Does aid for education educate children? Evidence from panel data. *World Bank Economic Review*, *22*(2), 291–314. https://doi.org/10.1093/wber/lhn003

Mahalik, M. K., Villanthenkodath, M. A., Mallick, H., & Gupta, M. (2021). Assessing the effectiveness of total foreign aid and foreign energy aid inflows on environmental quality in India. Energy Policy, 149. https://doi.org/10.1016/j.enpol.2020.112015

Mawejje, J., & Odhiambo, N. M. (2020). The determinants of fiscal deficits: a survey of literature. *International Review of Economics*, *67*(3), 403–417. https://doi.org/10.1007/s12232-020-00348-8

McCarthy, C., Sternberg, T., & Banda, L. B. (2023). Does Environmental Aid Make a Difference? Analyzing Its Impact in Developing Countries. *Land*, *12*(10). https://doi.org/10.3390/land12101953

Mohanty, R. K. (2020). Fiscal Deficit and Economic Growth Nexus in India: A Simultaneous Error Correction Approach. *Journal of Quantitative Economics*, *18*(3), 683–707. https://doi.org/10.1007/s40953-020-00211-1

Mohanty, R. K. (2018). *Fiscal Deficit and Economic Growth Linkage in India: Impact of FRBM Act* (pp. 89–105). https://doi.org/10.1007/978-981-10-6217-9_8

Mundle, S., Bhanumurthy, N. R., & Das, S. (2011). Fiscal consolidation with high growth: A policy simulation model for India. *Economic Modelling*, *28*(6), 2657–2668. https://doi.org/10.1016/j.econmod.2011.08.001

Niño-Zarazúa, M. (2016). Aid, education policy, and development. *International Journal of Educational Development*, *48*(48), 1–8. https://doi.org/10.1016/j.ijedudev.2015.12.002

Parjan, Ekohariadi, Suhartini, R., & Anistyasari, Y. (2025). The Effect of Self-Efficacy and Problem-Solving Skills on Social Skills through On-The-Job Training in Indonesian Aviation Cadets. *Journal of Ecohumanism*, *4*(1), 4133–4159. https://doi.org/10.62754/joe.v4i1.6300

Patra, B., & Sethi, N. (2024). Financial development and growth nexus in Asian countries: mediating role of FDI, foreign aid and trade. *International Journal of Social Economics*, *51*(5), 623–640. https://doi.org/10.1108/IJSE-09-2022-0587

Pradhan, K. (2020). *GROWTH-MAXIMISING FISCAL RULE TARGETS IN INDIA* (238). www.mids.ac.in

Sasmal, J., & Sasmal, R. (2020). Public Debt, Economic Growth and Fiscal Balance: Alternative Measures of Sustainability in the Indian Context. *Global Business Review*, *21*(3), 780–799. https://doi.org/10.1177/0972150918778940

Sethi, D., Mohanty, A. R., & Mohanty, A. (2020). Has FRBM rule influenced fiscal deficit-growth nexus differently in India? *Macroeconomics and Finance in Emerging Market Economies*, *13*(1), 53–66. https://doi.org/10.1080/17520843.2019.1677736

Sethi, N., Bhujabal, P., Das, A., & Sucharita, S. (2019). Foreign aid and growth nexus: Empirical evidence from India and Sri Lanka. *Economic Analysis and Policy*, *64*, 1–12. https://doi.org/10.1016/j.eap.2019.07.002

Tung, L. T. (2018). The effect of fiscal deficit on economic growth in an emerging economy: Evidence from Vietnam. *Journal of International Studies*, *11*(3), 191–203. https://doi.org/10.14254/2071-8330.2018/11-3/16

Özkan, O., Destek, M. A., Balsalobre-Lorente, D., & Esmaeili, P. (2024). Unlocking the impact of international financial support to infrastructure, energy efficiency, and ICT on CO2 emissions in India. *Energy Policy*, *194*. https://doi.org/10.1016/j.enpol.2024.114340